\documentclass[12pt, onecolumn]{revtex4}    % Font size (12pt) and column number (one or two).

\usepackage[a4paper, left=2.5cm, right=2.5cm,
 top=2.5cm, bottom=2.5cm]{geometry}       % Defines paper size and margin length

\usepackage[font=small,labelfont=bf]{caption}   % Defines caption font size and caption title bolded

\usepackage{dirtytalk}

\usepackage{subcaption}
\usepackage{graphics,graphicx,epsfig,ulem}	% Makes sure all graphics works
\usepackage{amsmath} 						% Adds mathematical features for equations
\usepackage{mathtools}
\usepackage{dsfont}

\usepackage{float}
\usepackage{setspace}

\usepackage{etoolbox}

\makeatletter
\patchcmd{\frontmatter@RRAP@format}{(}{}{}{}
\patchcmd{\frontmatter@RRAP@format}{)}{}{}{}
\renewcommand\Dated@name{}
\makeatother

\usepackage{fancyhdr}

\usepackage{xcolor}
\usepackage{braket}

\usepackage{enumitem}

\usepackage{hyperref}
\hypersetup{
    colorlinks=true,
    linkcolor=blue,
    filecolor=magenta,      
    urlcolor=cyan,
}

\makeatletter
\def\l@subsubsection#1#2{}
\makeatother

\begin{document}

\title{The quantum jump method: photon statistics and macroscopic quantum jumps of two interacting atoms} 
\date{Submitted: \today{}}
\author{Charles A. McDermott}
\affiliation{ Department of Physics, Durham University, Durham, DH1 3LE, UK}

\begin{abstract}              
\noindent
We first use the quantum method to replicate the well-known results of a single atom relaxing, whilst
demonstrating the intuitive picture it provides for dissipative dynamics. By use of individual ``quantum trajectories'', the method allows for simulation of systems inaccessible to ensemble treatments. This is shown by replicating resonance fluorescence, allowing us to concurrently demonstrate the method's facilitation of calculating photon statistics by the creation of discrete photon streams. To analyse these,
we solidify the theoretical basis for, and implement, a computational method of calculating second-order coherence functions. 
A process by which to model interacting two-atom systems to allow for computation with the quantum jump method is then developed. 
Using this, we demonstrate cooperative effects leading to greatly modified emission spectra, before investigating the decoupling of states from dissipative and coherent interactions. Here, we find the novel insight provided by the quantum jump method both births and provides the tools with which to begin an investigation into the occurrence of macroscopic jumps and the formation of macroscopic dark periods in a system of two two-level dipole-dipole coupled atoms.

\end{abstract}

\maketitle
\begin{figure}[b]
    \centering
    \includegraphics[width=0.4\columnwidth]{"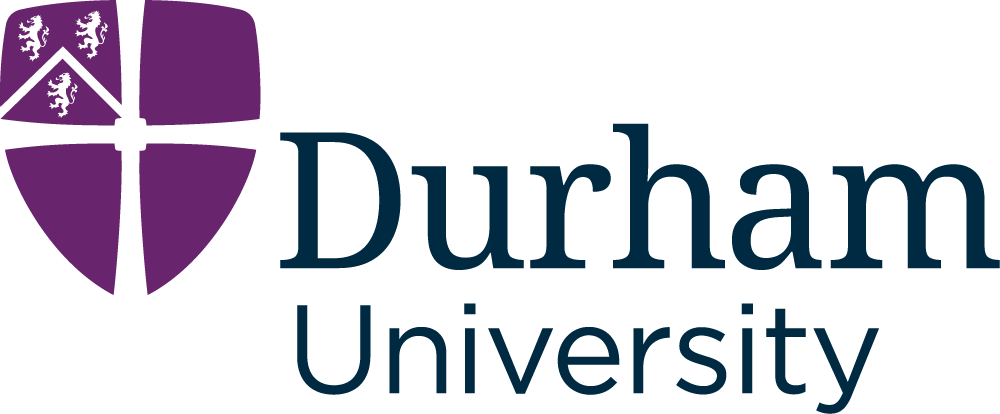"}
\end{figure}
\newpage
{\setstretch{1.1}
\tableofcontents}
\let\toc@pre\relax
\let\toc@post\relax

\newpage

\section{Introduction} 
\label{sec:Introduction}

{\setstretch{1.11}
For over a half-century since its inception, quantum mechanics served as a statistical theory making probabilistic predictions of the behaviour of ensembles and it was widely believed to be impossible to manipulate single quantum systems. Schrodinger himself stated that ``we are not experimenting with single particles, any more than we can raise Ichthyosauria in the zoo.''\cite{GK} Indeed, the idea of instantaneous transitions between states, first postulated by Bohr,\cite{Bohr} appeared in acute conflict with the continuity of wave mechanics and did not survive the field's revolution in the 1920s. However, the concept of quantum jumps did receive a breath of life in the interpretation of state collapse during measurement processes,\cite{Revival} and, in the 1950s, pioneering developments by Dehmelt and others began to pave the way towards isolating single quantum systems, challenging the previous ensemble status-quo.\cite{Traps} 

The ability to isolate small quantum systems required the development of theoretical frameworks capable of describing open-system dissipative dynamics, as no systems in a practical laboratory setting are perfectly isolated leading to coherent dynamics as described by the Schrödinger equation lasting only short timescales before becoming dominated by coupling to the environment.\cite{ShortT} However, the concept of quantum jumps was overshadowed by the development of master equation approaches, deterministic rate equations describing the time evolution of density matrices.\cite{MasterEquations1,MasterEquations2} It was not until the electron shelving experiments of Itano et al. \cite{Shelving}  that quantum jumps re-entered the foray, as ensemble descriptions could not account for these sharp transitions. This renewed interest led to the concurrent development of conditional dynamics and quantum trajectory techniques by multiple groups,\cite{MCWF,Dum1,Dum2,Photodetection,Gardiner,HegerfeldtQJM} consequently showing quantum jumps to be implicit in any standard photodetection theory.\cite{Photodetection}

It is the development of the ``Monte-Carlo Wavefunction method"  by Mølmer et al. on which this report is based.\cite{MCWF} Initially developed as a numerically efficient way to simulate laser cooling and intermediary sized quantum systems, the method utilises conditional evolution of a pure-state wavefunction. In the few-body systems we consider in this report, the quantum jump method's power lies in its ability to simulate individual ``quantum trajectories''. These individual trajectories serve to provide deeper insight to accompany ensemble treatments, and by recording jump times can produce discrete photon streams for statistical analysis.  As we will go on to show, this allows us to replicate known results, and both demonstrate and further investigate striking system dynamics arising from seemingly simple systems.

We begin in Chapter 2 with a brief overview of the method's theoretical basis, followed by the method itself in Chapter 3, and a simple, yet insightful demonstration in Chapter 4. Next, Chapter 5 investigates the semi-classical driving of a single atom, providing the components for modelling our more complex systems, and allowing us to introduce discrete photon statistics. In Chapter 6 we marry multiple works to provide a method to model two-body, coupled systems using the quantum jump method, and in Chapter 7 we use these methods to demonstrate known, dipole-dipole induced phenomena. Chapter 8 then moves on to investigate the fascinating phenomena of macroscopic dark periods in the emission characteristics of these two-atom systems, followed finally by our aims for further work and closing remarks in Chapters 9 \& 10.

\newpage
\section{Theory}
\label{sec:Theory}

\subsection{Open Quantum Systems \& Core Approximations}
\label{subsec:OpenQuantumSystems}
\vspace{-0.2cm}
%kj\subsubsection{Open Quantum Systems}

In order to describe dissipative dynamics and open systems, we begin by considering a total system that constitutes a closed quantum system by itself. This total system is divided into our system of interest, from now on referred to as the system, and an environment, with a dissipative interaction between the two characterised by the term $\Gamma$.\cite{OpenSystems}
As such, the total Hamiltonian is given by
\begin{equation}
    H_{T}=H_{S}\otimes \mathds{1}_{E} +\mathds{1}_{S} \otimes H_{E}+H_{I},
\label{eq:totalHamiltonian}
\end{equation}
where $H_{E}$ is the Hamiltonian describing the dynamics of the environment's degrees of freedom, $H_{I}$ describes the coupling between the environment and our system of interest, and $H_{S}$ is the Hamiltonian of our system of interest, describing the coherent dynamics of solely the system's degrees of freedom in the absence of any coupling to the environment. 

\begin{figure}[h!]
    \centering
    \includegraphics[width=0.5\textwidth]{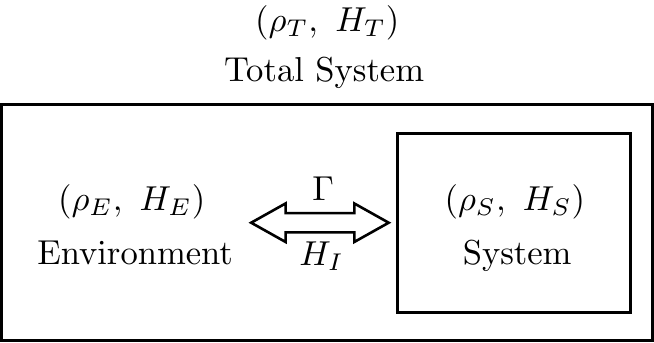}
    \caption{Schematic of an open quantum system: a closed total system comprised of a system coupled to an environment, with the interaction strength characterised by the parameter $\Gamma$.}
    \label{fig:opensystems}
\end{figure}

%\subsubsection{Core Approximations}
\label{subsec:CoreApproximations}

An essential aspect as to why we can describe the systems in quantum optics is the fact that we are able to make key assumptions that are in general difficult to justify for other systems.\cite{Approximations} For our systems of interest we make the following three key approximations in the interaction between the system and environment:

\begin{enumerate}[leftmargin=6mm]
    \item \textit{The Rotating Wave approximation}: By transforming into the reference frame rotating with the frequencies of the system and environment, (known as the interaction picture), non-energy conserving terms in the interaction Hamiltonian oscillate much faster than the typical timescale of the system's evolution. In this case, the effects of these terms will average to zero over the relevant timescales of the dynamics of our system, and we can neglect them.
    
    \item \textit{The Born approximation}: As the frequency scales associated with the coupling induced dynamics between the environment and system are small in comparison to that of both the frequency scales of our system of interest and the environment respectively, we are able to make a Born approximation in our derivation of the master equation of the system.
    
    \item \textit{The Markov approximation}: We assume that the system-environment coupling is local in time, that is $\dot{\rho}_{S}$ depends only on $\rho_{S}$ at the same time. This assumes that the environment is in effect memory-less and returns rapidly to equilibrium, unaffected by its coupling to the system, and allows us to set initial conditions. 
    
\end{enumerate}
These approximations rely on characteristic frequency scales being many orders of magnitude larger than $\Gamma$, and for typical optical systems this ratio is roughly $1\times10^{7}$, such that these approximations hold true to many orders of magnitude.\cite{Andrew}

\subsection{The Lindblad Master Equation}
\label{subsec:TheLindbladMasterEquation}
\vspace{-0.2cm}

As we treat the total system as a closed quantum system, we can describe the evolution of its density matrix with the Von-Neumann equation

\begin{equation}
    \dot{\rho}_{T}=-\frac{i}{\hbar}[H_{T},\rho_{T}(t)]\equiv \mathcal{L}(\rho),
\label{eq:vonneumann}
\end{equation}
where $\mathcal{L}$ is a so called Liouvillian superoperator. This superoperator acts on elements Fock-Liouville Hilbert space, a Hilbert space of density matrices with a defined scalar product. \cite{Manzano}
As we are interested in the behaviour and dynamics of solely the system, we obtain the reduced system density operator $\rho_{S}$ by tracing the total density operator over the environment degrees of freedom, giving $\rho_{S} =$ Tr$_{E} \{ \rho_{T} \}$. 
From this, through an extensive series of calculations using the aforementioned approximations, one obtains the equations of motion for the system degrees of freedom as
\begin{equation}
\dot{\rho}_{S} = \frac{i}{\hbar}[\rho_{S}, H_{S}] + \mathcal{L}_{\text{relax}}(\rho_{S}).
\label{eq:generalmasterequation}
\end{equation}
Here, the commutator describes the coherent evolution of our system in the absence of dissipative coupling, and the term $\mathcal{L}_{\text{relax}}$ is a superoperator acting on the reduced density matrix, governing the dissipative processes that arise due to the system's coupling to the environment.

In this report, we deal with atoms and assume that environment with which our systems couple to are ordinary vacuums, and that the bandwidths of these vacuum fields are broadband. Thus, the relaxation superoperator takes the general form:
\begin{equation}
\mathcal{L}_{\text{relax}} = -\frac{1}{2} \sum_{i,j}\gamma_{ij}(S_{i}^{+} S_{j}^{-} \rho_{S} + \rho_{S}  S_{i}^{+} S_{j}^{-} -2 S_{j}^{-} \rho_{S} S_{i}^{+} ),
\label{eq:relaxationsuperoperator}
\end{equation}
where each $\gamma_{ij}$ is the collective decay constant, and $S^{\pm}_{m}$ are dipole raising and lowering operators.\cite{Ficek} This results in the following markovian master equation
\begin{equation}
    \dot{\rho}_{S} = \frac{i}{\hbar}[\rho_{S}, H_{S}]  -\frac{1}{2} \sum_{i,j}\gamma_{ij}(S_{i}^{+} S_{j}^{-} \rho_{S} + \rho_{S}  S_{i}^{+} S_{j}^{-} -2 S_{j}^{-} \rho_{S} S_{i}^{+} ),
    \label{eq:lindblad}
\end{equation}
which is known as the Lindblad master equation. The importance of this equation cannot be overstated, with key roles in quantum optics, condensed matter physics, atomic physics and quantum biology to name but a few.\cite{Condensed, Atomic, Biology} Full derivations are astoundingly heavy, however an exceptional introduction and walk-through of the process is given in Ref.\cite{Manzano}.

\section{The Quantum Jump Method}
\label{sec:TheQuantumJumpMethod}

\subsection{The Effective Hamiltonian}
\label{subsec:TheEffectiveHamiltonian}
\vspace{-0.2cm}
The essence of the quantum jump method of Mølmer et al.\cite{MCWF} lies in the conditional evolution of the pure-state system wavefunction to form a trajectory, followed by averaging over the ensemble of all possible trajectories. To do this, we make use of a diagonalised form of the relaxation superoperator where there are no cross-terms, thus allowing us to sum over a single index $m$ in Eq.(\ref{eq:lindblad}). The superoperator is written in terms of Lindblad or ``jump'' operators, $C_{m}$, and takes the following form:
\begin{equation}
\mathcal{L}_{\text{relax}} = -\frac{1}{2} \sum_{m}(C_{m}^{\dagger} C_{m} \rho_{S} + \rho_{S}  C_{m}^{\dagger} C_{m}) + \sum_{m} C_{m} \rho_{S} C_{m}^{\dagger} .
\label{eq:mcwfrelaxationsuperoperator}
\end{equation}
It is straightforward to compare Eqs. (\ref{eq:relaxationsuperoperator}) and (\ref{eq:mcwfrelaxationsuperoperator}) for the relaxation superoperator in order to determine the form of our jump operators. It is these jump operators specifically that allow us to implement the quantum jump method by defining an effective Hamiltonian 
\begin{equation}
H_{\text{eff}} = H_{S} - \frac{i\hbar}{2}\sum_{m}C^{\dagger}_{m}C_{m}.
\label{eq:nonHermitianHamiltonian}
\end{equation}
This Hamiltonian is non-unitary, and the term summing over the jump operators is often referred to as the recycling term  as it acts to recover the population lost from certain states due to the non-hermiticity of $H_{\text{eff}}$ and place it in other states.

\subsection{Simulating Trajectories}
\label{subsec:SimulatingTrajectories}
\vspace{-0.2cm}

\begin{figure}[b!]
    \centering
    \includegraphics[width=\linewidth]{"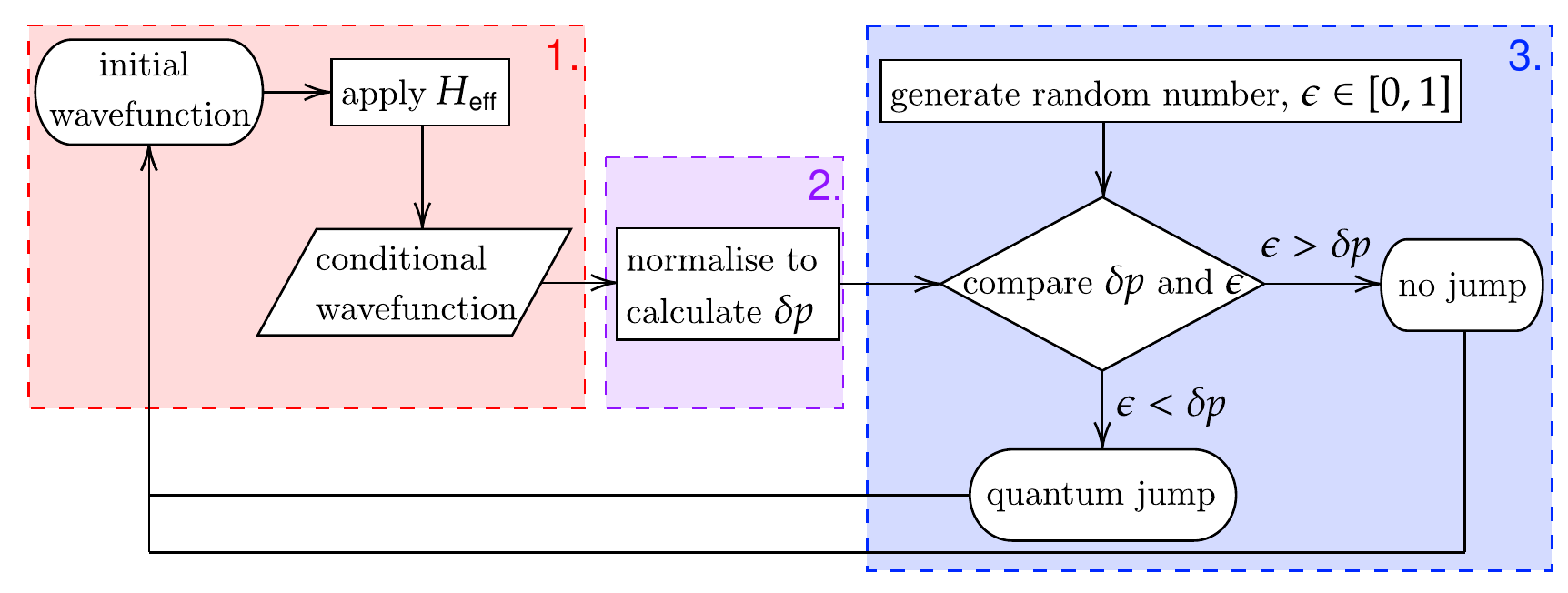"}
    \caption{Algorithm for the three parts of each time step of the quantum jump method.}
    \label{fig:qjmalgo}
\end{figure}

In order to simulate a single trajectory, the quantum jump method uses $H_{\text{eff}}$ and $C_{m}$ to evolve the wavefunction time steps $\delta t$, followed by re-normalisation and conditional checks to determine whether a quantum jump occurs. Outlined in Fig.\ref{fig:qjmalgo}, the algorithm to simulate a single trajectory consists of three main parts per time step. Step \color{red}1. \color{black}is to evolve the wavefunction through a time step, $\delta t$, using $H_{\text{eff}}$ to obtain a conditional wavefunction. Here, we assume that at some time $t$, the isolated system state is given by the normalised wavefunction $\ket{\psi (t)}$. For sufficiently small $\delta t$, the conditional wavefunction, $\ket{\psi^{(1)}(t + \delta t)}$, is given to first order by
\begin{equation}
\ket{\psi^{(1)}(t + \delta t)} = \left( 1 - \frac{iH_{\text{eff}} \, \delta t}{\hbar}\right)\ket{\psi(t)} . 
\label{eq:phi1}
\end{equation}
As this evolution is non-unitary, Step \color{violet}2. \color{black}is to take the square of the norm of the conditional wavefunction to obtain an associated renormalisation factor $\delta p$.
\begin{equation}
    \braket{\psi^{(1)}(t + \delta t)|\psi^{(1)}(t + \delta t)} = \bra{\psi(t)}\left( 1 + \frac{iH_{\text{eff}}^{\dagger} \, \delta t}{\hbar}\right) \left( 1 - \frac{iH_{\text{eff}}^{{\color{white}\dagger}} \, \delta t}{\hbar}\right)\ket{\psi(t)} = 1-\delta p ,
\label{eq:norm}
\end{equation}
where we find that $\delta p$ is given by
\begin{equation}
    \delta p = \delta t \frac{i}{\hbar}\bra{\psi(t)}H_{\text{eff}}-H_{\text{eff}}^{\dagger}\ket{\psi(t)} = \sum_{m}\delta p_{m},
\label{eq:dp}
\end{equation}
\begin{equation}
    \delta p_{m} = \delta t \bra{\psi(t)} C_{m}^{\dagger} C_{m} \ket{\psi(t)} \geq 0.
\label{eq:dpm}
\end{equation}
The magnitude of $\delta t$ is adjusted so that the above first order calculations are valid, requiring $\delta p \ll 1$.
Step {\color{blue}3.} is to then generate a random number, $\epsilon$, uniformly distributed between 0 and 1, and compare it to $\delta p$. If the randomly generated number is larger than $\delta p$, which occurs in most cases as $\delta p$ is much less than unity,  then the new wavefunction at time $t + \delta t$ is given by the normalised conditional wavefunction.
\begin{equation}
    \ket{\psi(t+\delta t)}=\frac{\ket{\psi ^{(1)}(t+\delta t)}}{(1-\delta p)^\frac{1}{2}}, \mathrlap{\qquad\quad\qquad \,\, \epsilon > \delta p.}
\label{eq:nojumpevolution}
\end{equation}
If in fact $\epsilon < \delta p$, then a quantum jump occurs, and we choose a new normalised wavefunction of a jump operated initial wavefunction as
\begin{equation}
    \ket{\psi(t+\delta t)}=\frac{C_{m}\ket{\psi(t)}}{(\delta p_{m}/\delta t)^\frac{1}{2}},\mathrlap{\qquad\qquad\qquad \epsilon < \delta p,}
\label{eq:quantumjumpevolution}
\end{equation}
where the respective jumped state is chosen according to the probability law $\prod_{m}= \delta p_{m}/\delta p$. The new normalised wavefunction is then used as the initial wavefunction for the next time step, and the above process is then repeated over the duration of the simulation. In this way, one defines a stochastic walk of the system wavefunction to produce a single quantum trajectory. It can be shown that, provided the initial state is normalised, by taking the ensemble average of all possible trajectories the quantum jump method is at all times equivalent to the equations of motion derived by a master equation treatment, regardless of the time step $\delta t$ chosen.\cite{MCWF} However, in order to avoid Zeno-effect-like pitfalls, one must choose $\delta t$ sufficiently larger than the coherence time of the environment.\cite{Zeno} This gives the rough requirement of $ \delta t \approx 10^{-13} - 10^{-10}$s.\cite{Revival} In general, we assume that system dynamics take place over the order of nanoseconds, and as such $\delta t$ is chosen as $0.1\Gamma$ for quick simulations, or $0.001\Gamma$ for detailed state dynamics.

\subsection{Measurement Theory and Physical Interpretation}
\label{subsec:MeasurementTheory}
\vspace{-0.2cm}

So far, we have outlined the quantum jump method as a computational process, however, the method carries links to quantum mechanics in a much more fundamental sense. As we treat atomic systems and write Eq.(\ref{eq:lindblad}) in terms of dipole operators, the quantum jump method carries a direct equivalence to choice of measurement protocol, and this link is completely intrinsic, arising from the theoretical basis for the method; the choice to unravel Eq.(\ref{eq:generalmasterequation}) in terms of jump operators $C_{m}$ to obtain Eq.(\ref{eq:mcwfrelaxationsuperoperator}) is equivalent to choosing to conduct measurements with $N$ ideal photodetectors over all space, where $N$ is the maximum index.\cite{IdealDetectors} This equivalence goes both ways, with choosing to model quantum jumps corresponding to a standard photodetection scheme, and, as shown by Carmichael, a standard photodetection scheme corresponding to quantum jumps.\cite{Photodetection}
Thus, for all atomic systems considered in this report, the jump operators correspond to a photodetection event, and our quantum jumps are equivalent to spontaneous emission events. 

In fact, the quantum jump method falls within the broad framework of continuous measurement theory.\cite{Dum1} Thus, each time step in our simulations corresponds to us conducting a gedanken broadband photon measurement in a window of time $\delta t$, over all space and with ideal detectors.\cite{ContMeasure} It should be noted that the correspondence of jumps to such measurable properties is not the case for all systems, such as the example of two-level relaxation with dephasing given by Mølmer et al.\cite{MCWF} Thus, in general one must be wary before prescribing physical meaning to the jumps. Remarkably, modern experimental techniques allow for tight enough monitoring of small open quantum systems so as for the above discussion to be more than just idealised.\cite{QuickMeasurement}

\subsection{Benefits}
\label{subsec:Benefits}
\vspace{-0.2cm}
In this report, the equivalence of quantum jumps and emission events for atomic systems provides the method great usefulness in our investigations of quantum optics. As noted in the previous section, to recover ensemble average descriptions we average this process over many trajectories. However, the power of the method when investigating few-body dissipative dynamics lies in the ability to simulate a single system by running an individual trajectory. Alongside allowing for the generation of discrete photon streams that can be statistically analysed, this provides state evolution that in some cases offers physical insight above and beyond that of master equation approaches.

Whilst we investigate few-body systems, the method was developed to offer a significant advantage for numerically simulating intermediary sized systems. It propagates a wavefunction as opposed to a density matrix, thus, for systems with relevant Hilbert space size $N$, the number of variables involved in calculation is of the order $\simeq N$ as opposed to $\simeq N^2$. However, as this advantage is only maintained if the number of trajectories averaged over is less than the size of the Hilbert space, it poses little benefit to this report where the number of trajectories are of similar order to or much greater than $N$.

\section{A Single Atom Relaxing} 
\label{sec:SingleAtomRelaxing}

\subsection{Modelling}
\label{subsec:SAModelling}
\vspace{-0.2cm}

The evolution of a single atom coupled to a continuum of electric field modes was solved in 1930 by Weisskopf and Wigner, who derived the exponential decay of spontaneous emission without requiring the hypothesis of quantum jumps.\cite{Wigner} However, it provides a simple example with which to demonstrate the implementation and features of the quantum jump method.
Simple rate equation treatments modelling damping of the atom yield the well known exponential decay of population in the excited state, giving:
\begin{equation}
    \dot{\rho}_{ee}=-\Gamma\rho_{ee},
\end{equation}
where $\rho_{ee}$ is the population of the excited state and $\Gamma$ is the lifetime of this state.\cite{GK} To implement the quantum jump method, we start by modelling a single two-level atom, setting the zero energy to be halfway between the excited and the ground state. The Hamiltonian of the system is given by,
\begin{equation}
    H_{S}=\frac{\hbar\omega}{2}\begin{pmatrix}
    1 & 0\\
    0 & -1
    \end{pmatrix},
\end{equation}
where $\omega$ is the atomic transition frequency. \textit{\underline{Note}: Throughout this report, we adopt a basis of the form $\{\ket{e}, ...,\ket{g}\}$, ordering the basis states from most to least energetic. All matrix representations of operators used will be given explicitly in the appendix for clarity.}

The system consists of a single atom and we assume a stable ground state. Thus, the relaxation superoperator sums over a single value of a single index, and there is single jump operator given by $C = \sqrt{\Gamma} S^{-}$, reducing Eq.(\ref{eq:relaxationsuperoperator}) to:
\begin{equation}
\mathcal{L}_{\text{relax}} = -\frac{1}{2}\Gamma(S^{+} S^{-} \rho_{S} + \rho_{S}  S^{+} S^{-} -2 S^{-} \rho_{S} S^{+} ),
\label{singleatomrelaxationsuperoperator}
\end{equation}
where $\Gamma$ is the lifetime of the excited state, arising from the atom's coupling to the environment, and $S^{+}$, $S^{-}$ are respectively the dipole raising and lowering operators. As such, our effective Hamiltonian given by Eq.(\ref{eq:nonHermitianHamiltonian}) takes the form,
\begin{equation}
    H_{\text{eff}} = H_{S} - \frac{i\hbar}{2}\Gamma S^{+}S^{-}.
\label{eq:Heffrelaxingatom}
\end{equation}
At a time $t=0$, we consider the atom to be in a superposition of the excited and ground states:
\begin{equation}
    \ket{\psi(0)}=\alpha_{0}\ket{e} + \beta_{0}\ket{g},
    \label{eq:phi0}
\end{equation}
where $\ket{e}$ denotes the excited state and $\ket{g}$ denotes the ground.
Propagating this inital wavefunction by use of Eq.(\ref{eq:phi1}), we obtain the unnormalised, conditional wavefunction as:
\begin{equation}
    \ket{\psi^{(1)}(\delta t)}=\alpha_{0} \left[ 1-\frac{i \omega \delta t }{2}- \frac{\Gamma \delta t}{2} \right]\ket{e} +\beta_{0} \left[1+\frac{i \omega \delta t }{2} \right]  \ket{g}.
\label{eq:phi1relaxingatom}
\end{equation}
The norm can then be calculated either numerically, or by use of Eq.(\ref{eq:norm}), and from Eq.(\ref{eq:dp}) we calculate the probability of making a quantum jump at each time step $\delta t$ as
\begin{equation}
    \delta p = \Gamma |\alpha_{0}|^{2} \delta t.
\end{equation}
This corresponds to the probability of emitting a photon between time 0 and $\delta t$, and thus, the probability of a spontaneous emission at any time is proportional to the occupation of the excited state at that time.

\subsection{Quantum Trajectories}
\label{sec:QuantumTrajectories}
\vspace{-0.2cm}

For the single atom system, self-written code was implemented and, for all simulations in this report, units are rescaled by a factor of $\Gamma$ to avoid computational errors arising from extreme values. Hence, simulations are evolved in time units of $\Gamma$ unless stated otherwise.

\begin{figure}[b!]
    \centering
    \includegraphics[width=\linewidth]{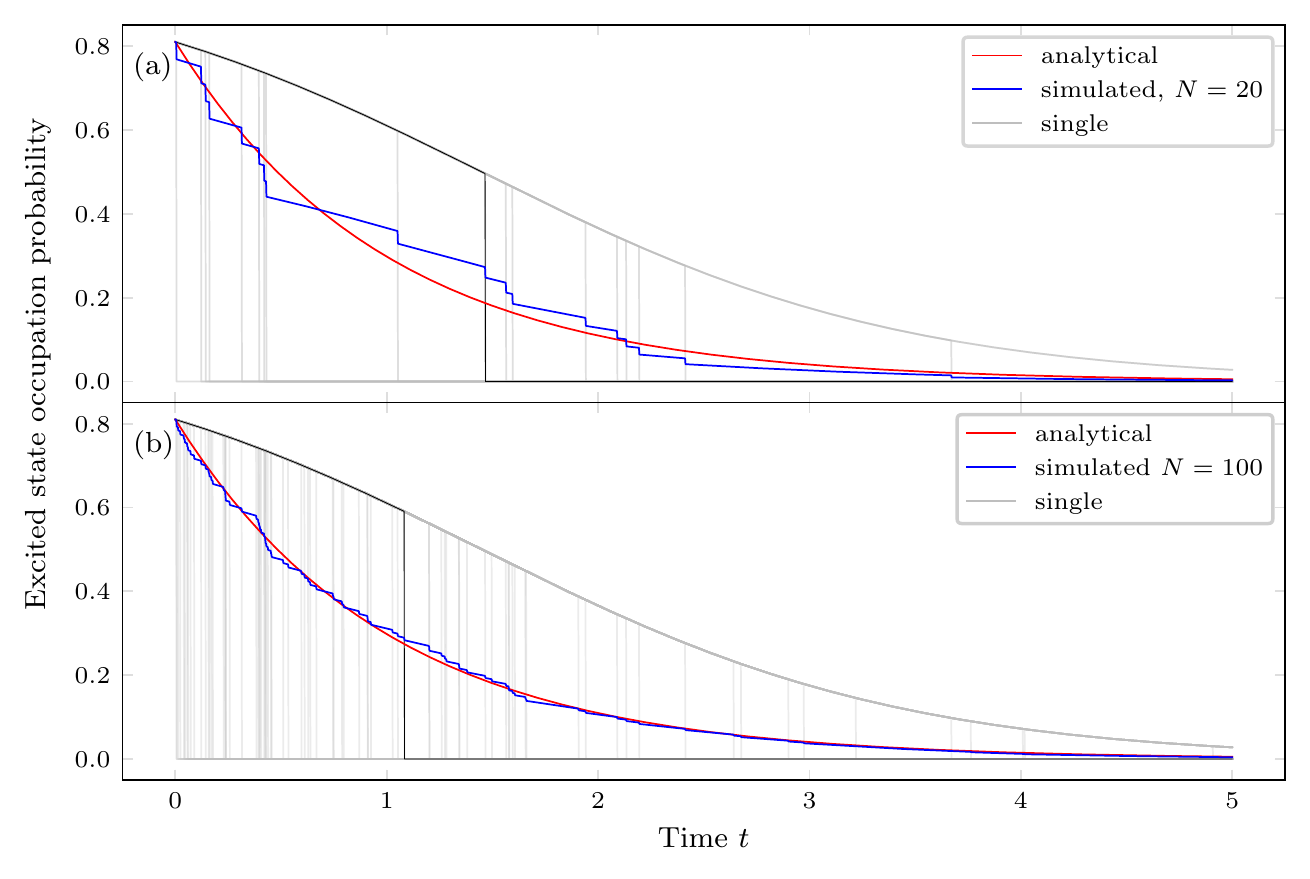}
    \vspace{-0.9cm}
    \caption{Evolution of excited state probability against time. Individual trajectories are plotted in grey, averaged trajectories for (a) N=20, (b) N=100, are plotted in blue, and the analytical prediction obtained by integrating the Optical Bloch Equations is plotted in red. A single trajectory in each figure is highlighted in black. The initial excited state coefficient was chosen as $\alpha_{0} = 0.9$, with $\beta_{0}$ such that the initial state had norm 1.  Simulations run with $\delta t = 0.001\Gamma$.}
    \label{fig:SA}
\end{figure}

Fig.\ref{fig:SA} shows the time evolution of the excited state occupation, $|\alpha|^{2}$, that we obtained from our simulations over two different ensemble sizes. As described in Section \ref{sec:TheQuantumJumpMethod}, each simulation is generated using multiple individual trajectories, and these can be seen in grey in Fig.\ref{fig:SA}, with each individual trajectory corresponding to simulating a single atom. The averaged trajectory in blue is then obtained by taking the ensemble average of the individual trajectories, and these simulated trajectories are then plotted against the result in red obtained from the optical Bloch equations.
We see that each individual trajectory undergoes a period of no jump evolution where it relaxes towards the ground state, interrupted at random points in time by a quantum jump, whereby the system is projected into the ground state and undergoes no further evolution. It is evident that simulations approach the analytical prediction for increasing ensemble size, and that we recover the results of well known theory. It is key to reiterate that it is the ensemble average of these individual trajectories, including both no jump evolution and quantum jumps, which recovers the results expected from the ensemble description.

%\subsection{Statistical Error in Averaged Trajectories}
%\subsubsection{Error}

\begin{figure}[h!]
    \centering
    \includegraphics[width=\linewidth]{"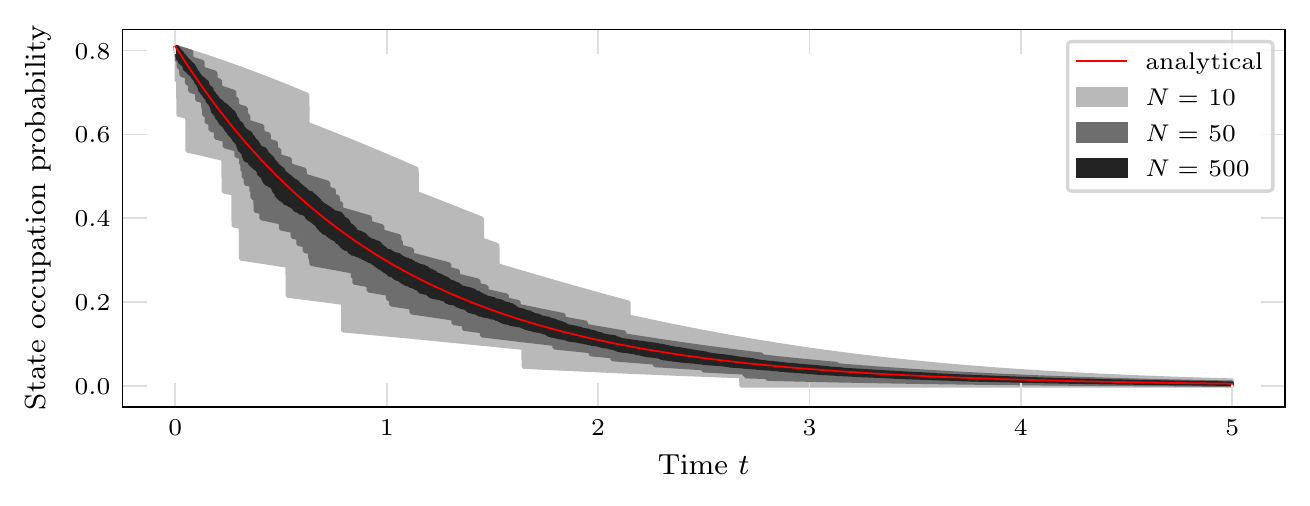"}
    \vspace{-0.8cm}
    \caption{Visual representation of the dispersion of simulations run with various numbers of particles. Each respective channel is created by generating 30 averaged trajectories at each respective N, then filling the space between the most extreme.}
    \vspace{-0.2cm}
    \label{fig:dispersion}
\end{figure}

As previously stated, we aim to implement the quantum jump method largely for its use as a conceptual device for visualising individual trajectories. However, we have generated Fig.\ref{fig:dispersion} to qualitatively illustrate to the reader the convergence of the quantum jump approach to master equation treatments. As shown by the widest channel in Fig.\ref{fig:dispersion}, for low $N$ it is possible to have largely deviating averaged trajectories. By increasing the number of atoms, and hence ensemble size, we observe a narrower channel of averaged trajectories, more closely tending to that of the theoretical fit, as expected. A practical write up of the method's statistical error can be found in Ref.\cite{Andrew}.

%\newpage
\subsection{No-Jump Evolution}
\label{subsec:NoJumpEvolution}
\vspace{-0.2cm}
From Fig.\ref{fig:SA} we observe that the ensemble is able to relax back to the ground state with only 18 quantum jumps, demonstrating one of the most interesting aspects of the quantum jump method; that of the no-jump evolution described by Eq.(\ref{eq:nojumpevolution}). 
To quantify this phenomena, we first note that whilst the unnormalised wavefunction in Eq.(\ref{eq:phi1relaxingatom}) allows us to simulate the evolution of the system computationally by simply renormalising numerically to calculate $\delta p$, additional insight may be gained by looking at the closed form of the normalised wavefunction. If a quantum jump does not occur, calculating the norm with Eq.(\ref{eq:dp}) and using the fact $\delta t$ is small, we obtain
\begin{equation}
    \ket{\psi(t)}=\alpha_{0}\left[1-\frac{\Gamma \delta t}{2} |\beta_{0}|^{2} \right]\text{exp} \left (-i \omega \delta t /2 \right) \ket{1} + \beta_{0} \left[1+\frac{\Gamma \delta t}{2} |\alpha_{0}|^{2} \right]\text{exp} \left (i \omega \delta t /2 \right) \ket{0}.
\end{equation}
In this form, we see a slight rotation of the wavefunction, in the square brackets, leading to an increase in probability of ground state occupation, and a concurrent decrease in probability of being in the excited state. As noted by Mølmer et al. this rotation is essential. Without it, the probability of photon detection (a quantum jump) over any two time intervals of width $\delta t$ would be equal, and one would incorrectly find that at some point between $t=0$ and infinity a photon would \textit{always} be emitted.\cite{MCWF} Instead, by undergoing no jump evolution, it is in fact possible for a single atom to relax to the ground state without ever emitting a photon.

It is here that the method's link to continuous measurement theory offers valuable insight into this phenomena of no-jump evolution. In the context of our single atom relaxing, whilst it is quite apparent that the detection of a photon updates our information of the system, consequently projecting it into the ground state, we importantly also gain information by measuring that no photon has been emitted, with the system subsequently being projected into the renormalised state evolved under the effective Hamiltonian; the observation of no photon informs us that the system is now somewhat more likely to be in the ground state. Thus, with the quantum jump method we are able to neatly and concisely demonstrate that a system's coupling to the environment affects the system's dynamics beyond simply the spontaneous emission of photons.

%\newpage
\section{Driven Single Atom}
\label{sec:DrivenSingleAtom}

\subsection{Modelling}
\label{subsec:DSAModelling}
\vspace{-0.2cm}

In this section, we consider the example of a single, two-level atom driven by laser light, and this will allow us to build key components for later systems, and introduce discrete photon statistics. We consider a semi-classical regime where we treat the laser field classically, and the atom quantum mechanically. The cases we consider are those of strong driving and, unlike perturbative methods, the quantum jump method allows full treatment of such systems. 
Under coherent semi-classical driving, the Hamiltonian of the atom can be shown to be given by the Rabi Hamiltonian \cite{QCQI}:
\begin{equation}
    H_{S} = H_{\text{opt}} =\frac{\hbar}{2}\begin{pmatrix}
        \Delta & \Omega\\[1mm]
        \Omega & -\Delta 
        \end{pmatrix}, \quad \Delta=\frac{1}{2}\left(\omega_{L} - \omega_{i} \right),
    \label{eq:Hopt}
\end{equation}
where $\Delta$ is the detuning of the frequency of the laser from the atomic transition frequency, and $\Omega$ is the Rabi frequency - proportional to the square root of field intensity and a measure of the strength of coupling between the system and the incident light. For our systems omega is taken as real and positive. As we deal again with a two-level system with stable ground state, the relaxation superoperator takes the same form as that of the single relaxing atom, given by Eq.(\ref{singleatomrelaxationsuperoperator}), yielding
\begin{equation}
    H_{\text{eff}}= H_{\text{opt}} - \frac{i\hbar}{2}\Gamma S^{+}S^{-},
    \label{eq:HeffDSA}
\end{equation}
as our effective Hamiltonian, seeing only a new form of the system Hamiltonian.

\subsection{Rabi Oscillations and Resonance Fluorescence}
\label{subsec:RabiOscillations}
\vspace{-0.2cm}

For a system evolved in time by the Rabi Hamiltonian we find that population oscillates between the excited and ground states. These oscillations are known as Rabi oscillations. Population is driven between the two states by the coherent laser field, with maximal population transfer limited by the detuning of the laser from the atomic transition frequency.\cite{QCQI} On resonance these oscillations take place at the Rabi frequency with complete transfer between the two states, however, for higher detunings the frequency of these oscillations decreases. We simulate this evolution by setting the decay rate $\Gamma$ to zero in Eq.(\ref{eq:HeffDSA}) and rescaling our variables by $1/\Omega$, obtaining the system evolution of Fig.\ref{fig:DSA}{\color{blue}(a)}.

\begin{figure}[b!]
    \centering
    %\hspace{-1cm}
    \includegraphics[width=\linewidth]{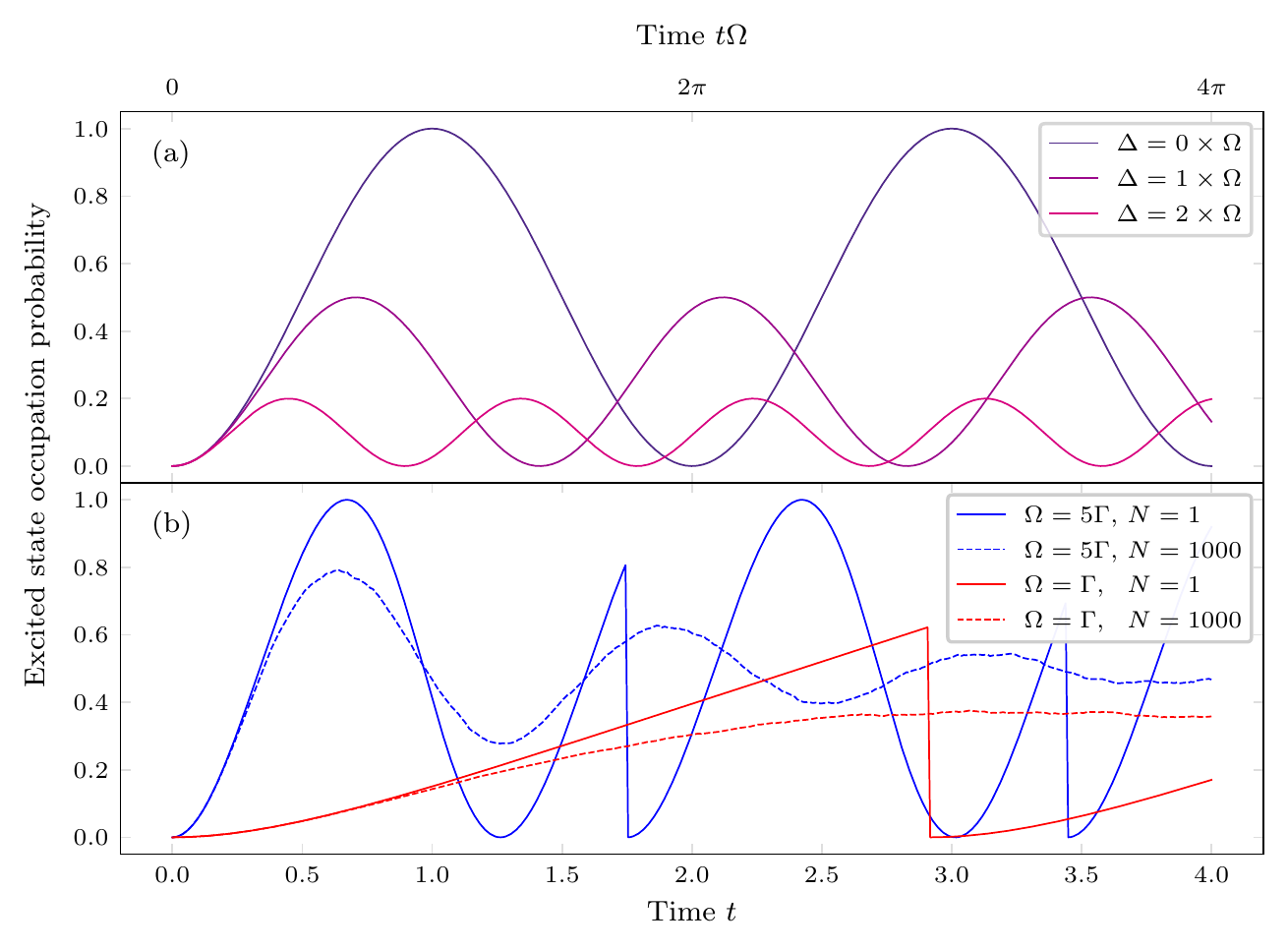}
    \vspace{-0.8cm}
    \caption{(a) Rabi oscillations of the excited state population. Simulations run with $\Gamma=0$, $\Omega=1$, $\delta t=0.001\Gamma$. with variables rescaled by $\Omega$, (b) Resonance fluorescence of a single atom. Simulations run with $\Gamma=1$, $\Delta=0$ and  $\delta t=0.001\Gamma$, for $\Omega=5\Gamma$ blue, and $\Omega=\Gamma$, red. }
    \vspace{-0.3cm}
    \label{fig:DSA}
\end{figure}

When reintroducing the dissipative coupling of the system to the environment, using methods describing the average behaviour of an ensemble of identical systems, such as the optical Bloch equations or averaging multiple trajectories, one only obtains the damped oscillations seen in dashed plots in Fig.\ref{fig:DSA}{\color{blue}(b)}. Instead, by simulating a single trajectory, the quantum jump method allows us to demonstrate the phenomena of resonance fluorescence.\cite{Scully} In this process, our two-level system is driven on resonance with its atomic transition frequency by a continuous coherent laser field, and undergoes periods of Rabi oscillation interrupted by spontaneous emission events projecting the system into the ground state. This is simulated by reintroducing, and again rescaling by, $\Gamma$, and we have chosen to simulate exactly on resonance with $\Delta = 0$. These results are shown in Fig.\ref{fig:DSA}{\color{blue}(b)}.

One may notice that the periods of oscillation in resonance fluorescence are in fact slightly slanted. This is easily understood as a consequence of the rotation of the wavefunction during no jump evolution that was discussed for the single atom relaxing. The rotation caused by the recycling term in $H_{\text{eff}}$ acts to transfer population from the excited, to the ground state, thus everywhere adding a negative contribution to the rate of change of population transfer into the excited state. This decreases the rate at which population is pumped into the excited state, whilst increasing the rate at which population is pumped from the excited state, thus introducing the slight slant.

%\newpage
\subsection{Second-order Coherence Functions}
\label{subsec:SecondOrderCoherenceFunctions}
\vspace{-0.2cm}

In this report, we investigate many systems fluorescing and emitting photons. As such, in order to determine the various statistical properties of the light emitted we make use of second-order coherence functions. These functions provide a way to determine the temporal coherence of a given light source, and were first introduced by Hanbury Brown and Twiss (HBT) as a way to measure stellar diameters in the 1950s.\cite{HBT} Classically, they were given in terms of time-averaged intensities, however, a quantum analogue to these functions was introduced by Glauber and others in the 1960s, closely resembling their classical counterparts but replacing intensities with products of electric field operators.\cite{Glauber} We specifically consider quantum second-order two-time coherence functions, denoted $g^{(2)}(\tau)$. These functions are proportional to the conditional probability that if a photon is detected at time $t$, one is also detected at time $t+ \tau$ \cite{GK}:
\begin{equation}
    g^{(2)}(\tau) \propto P(t+\tau |t),
\end{equation}
thus, giving us a way to quantify the statistical properties of discrete photon streams.

Qualitatively, there are in general three classifications of light when considering second-order discrete photon statistics, and these are depicted in Fig.\ref{subfig:lights}. In the case of antibunching, photons tend to arrive evenly spaced with suppressed correlations at $\tau=0$, quantified by $g^{(2)}(0)<g^{(2)}(\tau)$. Coherent light, such as that from a well-stabilised laser, exhibits zero correlation in photon arrival times, and as such, $g^{(2)}(\tau)$ is single valued and equal to 1 for all $\tau$. Finally, in bunched light, such as light from chaotic or thermal light sources, photons tend to arrive in clusters or bunches, yielding $g^{(2)}(0)>g^{(2)}(\tau)$.\cite{LightTypes}

\begin{figure}[t!]
    \centering
    \subfloat[]{\hspace{-3mm}\includegraphics[width=0.6\columnwidth]{"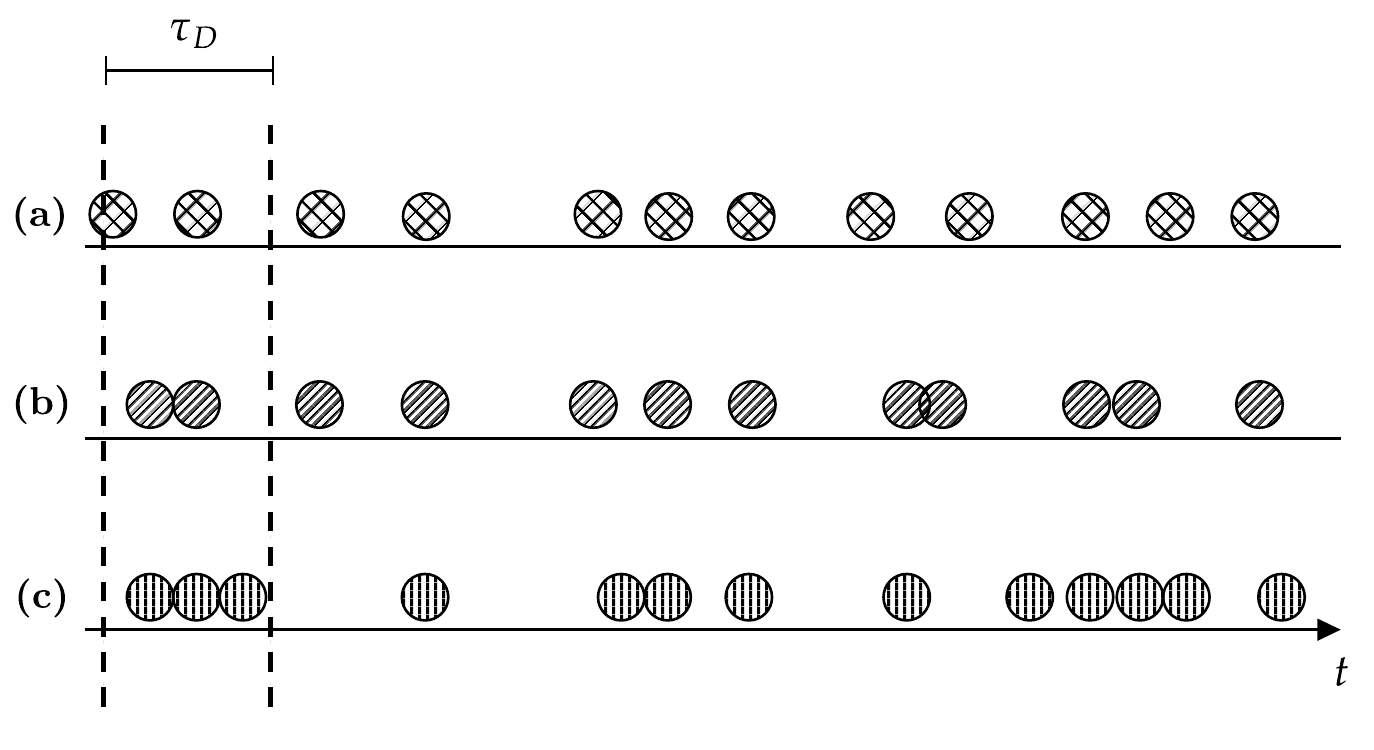"}%
        \label{subfig:lights}}
    \subfloat[]{\hspace{-3mm}\vspace{-5mm}\includegraphics[width=0.35\columnwidth]{"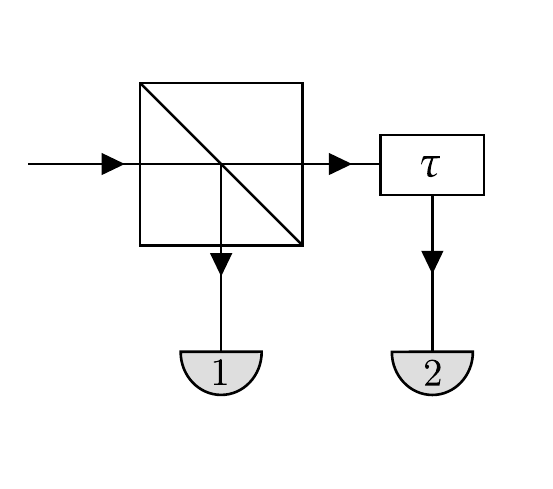"}%
    \hspace{-1cm}
        \label{subfig:HBT}}
    \caption{\textbf{(a)} A detector of fidelity $\Delta t_{D}$ measuring three streams of photon counts (a) Antibunched light where photons tend to be spaced apart with a low instance of multiple photon counts (b) Coherent (random) light where both the photon arrival times have no correlation (c) Bunched light whereby there is an increased instance in multiple photon detection.
    \textbf{(b)} HBT configuration interferometer composed of a 50:50 beam splitter and a time delay, $\tau$, on photons sent to detector 2.}
    \label{fig:photonstats}
\end{figure}

There are instances where the classical and quantum pictures agree. Descriptions of both coherent and bunched light using quantum coherence functions have classical analogues.\cite{ClassicalAgreement} However, there exists no classical analogue to describe antibunched light; a semi-classical coherence function would no longer carry the characteristic of a probability density, having seemingly negative probabilites. Thus, as $g^{(2)}(\tau)$ takes on classically forbidden values it gives us a way to identify purely quantum states of light.
There is strong motivation for classifying light sources producing these features; sources exhibiting antibunching characteristics finding applications in sensing, quantum computation and quantum cryptography, \cite{ABSensing, ABQComp, ABQCrypto}, and bunched light finds technological applications in interference experiments and imaging.\cite{BunchedInterference, BImaging} Thus, quantifying the characteristics of light sources plays a key role in many exciting fields at the forefront of modern physics.

\subsubsection{Deriving an equation for g2}

%\newpage
%\vspace{0.3cm}
%\begin{center}
 %   {
 %   \underline{\textit{Deriving $g^{(2)}(\tau)$}}}
%\end{center}
%\label{subsubsec:Derivingg2}
%\vspace{0.2cm}

Here, we derive a formula based on conditional probabilities using time-binned photon counts in order to obtain a method with which we can compute $g^{(2)}(\tau)$ using photon streams created from simulations.
To begin, we consider detectors in the HBT configuration as seen in Fig.\ref{subfig:HBT}. This is comprised of a lossless 50:50 beam splitter and two detectors, with the photon counts sent to Detector 2 undergoing a time delay $\tau$. There is a minor difference between correlation functions based on electric field operators and number operators, which we elaborate on in \ref{app:Electricfieldvs}, but for simplicity, we consider our detectors to be broadband and over all space.
These idealised detectors have a fidelity of $\Delta t_{D}$, corresponding to our time bin-width, and hence $g^{(2)}(\tau)$ is proportional to the probability distribution $p(n,\Delta t_{D})$ to detect a $n$ photons in the temporal interval $\Delta t_{D}$.\cite{G2propProbDist} 
In terms of detection probabilities, we obtain
\begin{equation}
    g^{(2)}(\tau)=\frac{P\left( D_{1} \text{ and } D_{2} \right)}{P\left(D_{1}\right)P\left(D_{2}\right)},
\end{equation}
where $P()$ denotes probability, $D_{1}$ is a photon detection at Detector 1 at time $t$, and $D_{2}$ is a photon detection at Detector 2 at time $t+\tau$.\cite{ProbG2}

To acquire these probabilities we follow a similar approach to Ref.\cite{BeigePhoton}. We begin by defining the number of photons detected in a certain time interval $(t_m, t_m+\Delta t_{D})$ at Detector 1 as $n_{1m}(t_m,t_m+\Delta t_{D})\equiv n_{1m}$, and at Detector 2 as $n_{2m}(t_m+\tau,t_m+\Delta t_{D}+\tau)\equiv n_{2m}^{\tau}$. Next, we consider a stream of photons produced by a single trajectory simulation sent through the HBT configuration interferometer. From time $t=0$ to time $t=T=M\Delta t_{D}$, the detectors measure whether or not a photon has been emitted in each time interval $(t_{m},t_{m+1})=(m\Delta t_D,(m+1)\Delta t_D)$. This allows us to calculate the relative frequency of cases in which a photon is detected at Detector 1 in the interval $(t_{m},t_{m+1})$ and at Detector 2 in the time-lagged interval $(t_{m} +\tau ,t_{m+1} +\tau)$ as
%$T\xrightarrow[]{} $ $\infty$ , $\Delta t_{D}\xrightarrow[]{} $ $0 $ 
\begin{equation}
    \begin{split}
       F(\tau) &=  \frac{1}{T} \sum_{m} n_{1}(t_{m},t_{m+1}) \times n_{2}(t_{m} +\tau ,t_{m+1} +\tau)\, \Delta t_{D} \\
       &= \frac{1}{T} \sum_{m} n_{1m} \times n_{2m}^{\tau}\, \Delta t_{D}, 
    \end{split}
    \label{eq:relfreq}
\end{equation}
where the summation is over all time bins. From this, we re-obtain the probability of coincident photon detection at detectors 1 and 2 in the following limit:
\begin{equation}
    P\left( D_{1} \text{ and } D_{2} \right) = \lim_{T\to\infty}\lim_{\Delta t_{D}\to 0} F(\tau).
\end{equation}
Employing the same approach to obtain $P(D_{1})$ and $P(D_{2})$, and using $T/\Delta t_{D} = M$ we obtain
\begin{equation}
 g^{(2)}(\tau)=\lim_{T\to\infty}\lim_{\Delta t_{D}\to 0}\frac{\sum_{m} n_{1m} \times n_{2m}^{\tau}}{\sum_{m} n_{1m} \sum_{m} n_{2m}^{\tau}} M,
 \label{eq:g2lim}
\end{equation}
where $M$ corresponds to the total number of measurements made by the detectors and hence the number of time-bins. 
It can be shown that this $g^{(2)}(\tau)$ is independent of detector efficiency, and that this single trajectory treatment is directly equivalent to that of an ensemble treatment in the limits of Eq.(\ref{eq:g2lim}). 
%Naturally we cannot take the complete limits, however, appropriate variable magnitudes are chosen. 

%\newpage

\subsubsection{Computational method to calculate $g^{(2)}(\tau)$ }
With the previous derivation, we provide the theoretical justification for a computational procedure to implement $g^{(2)}(\tau)$ that was outlined by Facão et al in Ref.\cite{Facao}.
We modify this procedure and outline it fully below, and find that by allowing for the recording of jump times in simulations, the quantum jump method coupled with Eq.(\ref{eq:g2lim}) provides a simple yet effective basis for numerically calculating photon statistics.

The algorithm to calculate $g^{2}(\tau)$ is as follows:
\begin{enumerate}
    \item Consider a stream of $N$ photons created by some process. Construct an array of counts, length $N$, with each element sequentially corresponding to the time that each respective photon in the stream was emitted:
    \begin{equation}
        C_{\text{Total}} = \left(t_{0}=0, t_{1}, t_{2}, ..., t_{N} \right),
    \end{equation}
    where the first photon's emission time, $t_{0}$, is treated as 0.
    
    \item Next, simulate a 50:50 beam splitter. Start by generating an array length $N$ of random numbers, with each element, $r_{i}$, uniformly distributed on the interval [0,1]. Now create two lists, $C_{1}$ and $C_{2}$, which will correspond to the detection times of photons arriving at detector 1 or detector 2 respectively. 
    For each element $t_{i}$ in $C_{\text{Total}}$, if $r _{i}<0.5$ then the corresponding photon goes to detector 1 and $t_{i}$ is appended to $C_{1}$. Visa versa for $r _{i}>0.5$.
    
    \item Now add the time delay $\tau$ to the photons sent to detector 2 by generating a new array, $C_{2}^{(\tau)}$, with each entry equal to $C_{2i}-\tau$. Truncate any negative time values.
    
    \item Account for the finite response time of each detector by creating time bins in which each photon will respectively be detected. The time response of each photodetector, which corresponds to the time bin width, should be sufficiently small such that it can resolve individual photons within the stream from one another. As such the response time, $\Delta t_{D}$ should be less than the average time between consecutive photons. 
    To bin the photons, calculate the number of time bins as 
    \begin{equation}
        n_{bins_i} = \left\lceil \frac{t_{max_i}}{\Delta t_{D}} \right\rceil ,
    \end{equation}
    where $\lceil x \rceil$ is the ceiling function and $t_{max_i}$ is the time of the final count in the stream being binned. Using $n_{bins_i}$, generate the bin edges and bin counts at both detectors using the Numpy histogram function to give us the arrays $N_{1}$ and $N_{2}^{\tau}$, with each element equal respectively to the number of photons detected in the corresponding time-bins at detector 1 and 2:
    \begin{equation}
        \begin{split}
            N_{1}=\left[ n_{11},\; n_{12},\; ...\; \right] \\
            N_{2}^{\tau}=\left[ n_{21}^{\tau},\; n_{22}^{\tau}, \; ... \; \right]
        \end{split}
    \end{equation}
    
    Next, truncate the arrays to be the same length. It is important to note that $t_{max}$ is now no longer the the time of the final photon count from the initial stream, $t_{N}$, due to the truncation of the arrays to be equal length.
    
    \item Now that we have the time-binned photon counts, calculate $g^{(2)}(\tau)$ in terms of discretised time averages of our count arrays:
    \begin{equation}
        g^{(2)}(\tau) =\frac{\braket{N_{1}N_{2}^{\tau}}_{t}}{\braket{N_{1}}_t \braket{N_{2}^{\tau}}_{t}},
        \label{eq:g2averages}
    \end{equation}
    with
    \begin{equation}
        \begin{split}
            &\braket{N_{1}N_{2}^{\tau}}_t  = \frac{1}{T} \sum_{i} n_{1i} n_{2i}^{\tau}\;\Delta t_{D} \;, \\
            &\braket{N_{1}}_t  = \frac{1}{T} \sum_{i} n_{1i} \;\Delta t_{D} \;, \quad
            \braket{N_{2}^{\tau}}_t  = \frac{1}{T} \sum_{i} n_{2i}^{\tau}\;\Delta t_{D} \;, \\ 
        \end{split}
    \label{eq:averages}
    \end{equation}
    where $T$ is the total time we are averaging over. As both arrays are the same length, we are averaging over the same time for both detectors, and $T$ corresponds to the time value of the final edge of our final time-bin: 
    \begin{equation}
        T = \text{len} \left( N_{2}^{\tau} \right) \times \Delta t_{D}.
    \end{equation}
    Thus, subbing Eqs.(\ref{eq:g2averages}) in to Eq.(\ref{eq:g2averages}) and cancelling, we arrive at the equation with which we calculate our second-order coherence function:
    \begin{equation}
        g^{(2)}(\tau) = \frac{\sum_i n_{1i} n_{2i}^{\tau}}{\sum_i n_{1i} \sum_i n_{2i}^{\tau}} \text{len} \left( N_{2}^{\tau} \right).
    \end{equation}
    This is equivalent to Eq.(\ref{eq:g2lim}) with $M=\text{len} \left( N_{2}^{\tau} \right)$ given that appropriate variable sizes are chosen to approach the given limits.
\end{enumerate}

\begin{figure}[h!]
    \centering
    \includegraphics[width=0.8\linewidth]{"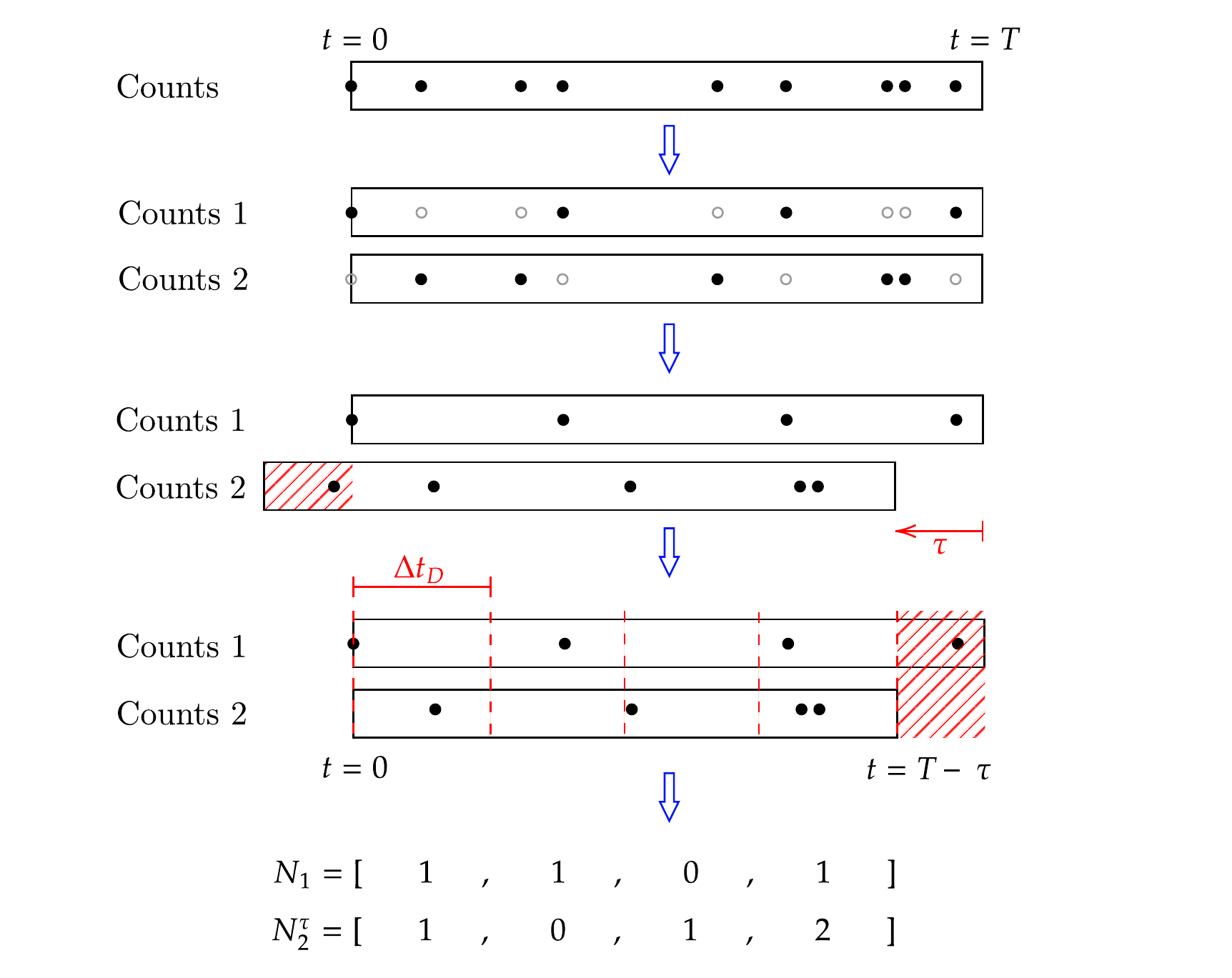"}
    \caption{Visualisation of the procedure used to generate the arrays of counts from which $g^{(2)}$ can be calculated using equationG2}
    \label{fig:computingg2}
\end{figure}

\subsection{Photon Statistics of Resonance Fluorescence}
\label{subsec:PhotonStatisticsofResonanceFluorescence}
\vspace{-0.2cm}

The previously discussed phenomena of antibunching was first predicted by Charmichael and Walls and was later observed by Kimble and Mandel in 1967 in the resonance fluorescence of single sodium atoms.\cite{Orszag} To produce this light, we again simulate resonance fluorescence using a single trajectory as in Fig.\ref{fig:DSA}{\color{blue}(b)}.

Fig.\ref{fig:g2s} shows the successful results of our simulations and subsequent computations of $g^{(2)}(\tau)$ of the light produced in various regimes of driving strength. In all cases, we observe suppressed correlations at $\tau=0$ as expected. As previously noted, this suppression which characterises antibunched light carries no classical analogue, but can be explained as follows.
Given a photon emitted at some time, this corresponds to our system being projected into to the ground state. As the atom now inhabits the lowest energy state, there is a finite time taken before the atom is driven back into an excited state where a photon may again be emitted. Thus, during this ``dark time'' there is a reduced probability of another photon being emitted with zero time delay. Hence, as $g^{(2)}(\tau)$ is proportional to the joint probability of coincident photon detection, we observe $g^{(2)}(0)<g^{(2)}(\tau)$.\cite{GK}

For weak driving regimes we observe a maximum value of $g^{(2)}(\tau)$  at $\tau \approx \tau _{P}$, where $\tau _{P}$ is the average time between consecutive photons. Further to this, as the driving strength is increased one also observes increasing Rabi oscillations in Fig.\ref{fig:g2s}{\color{blue}(b)} and {\color{blue}(c)}. For these strongly driven systems we observe the largest peaks in $g^{(2)}(\tau)$  at $\tau$ of the order $1/\Omega$, corresponding to enhanced emission probabilities due the system being strongly driven back into the excited state after a spontaneous emission event. In all driving regimes, we observe that for $\tau \gg \tau_{P}$,  $g^{(2)}(\tau)\to 1$ as the coherence time for the system, $\tau_{S} \sim \tau_{P}$, is exceeded.

\begin{figure}[b!]
    \centering
    %\hspace{-5mm}
    \vspace{-0.4cm}
    \includegraphics[width=\linewidth]{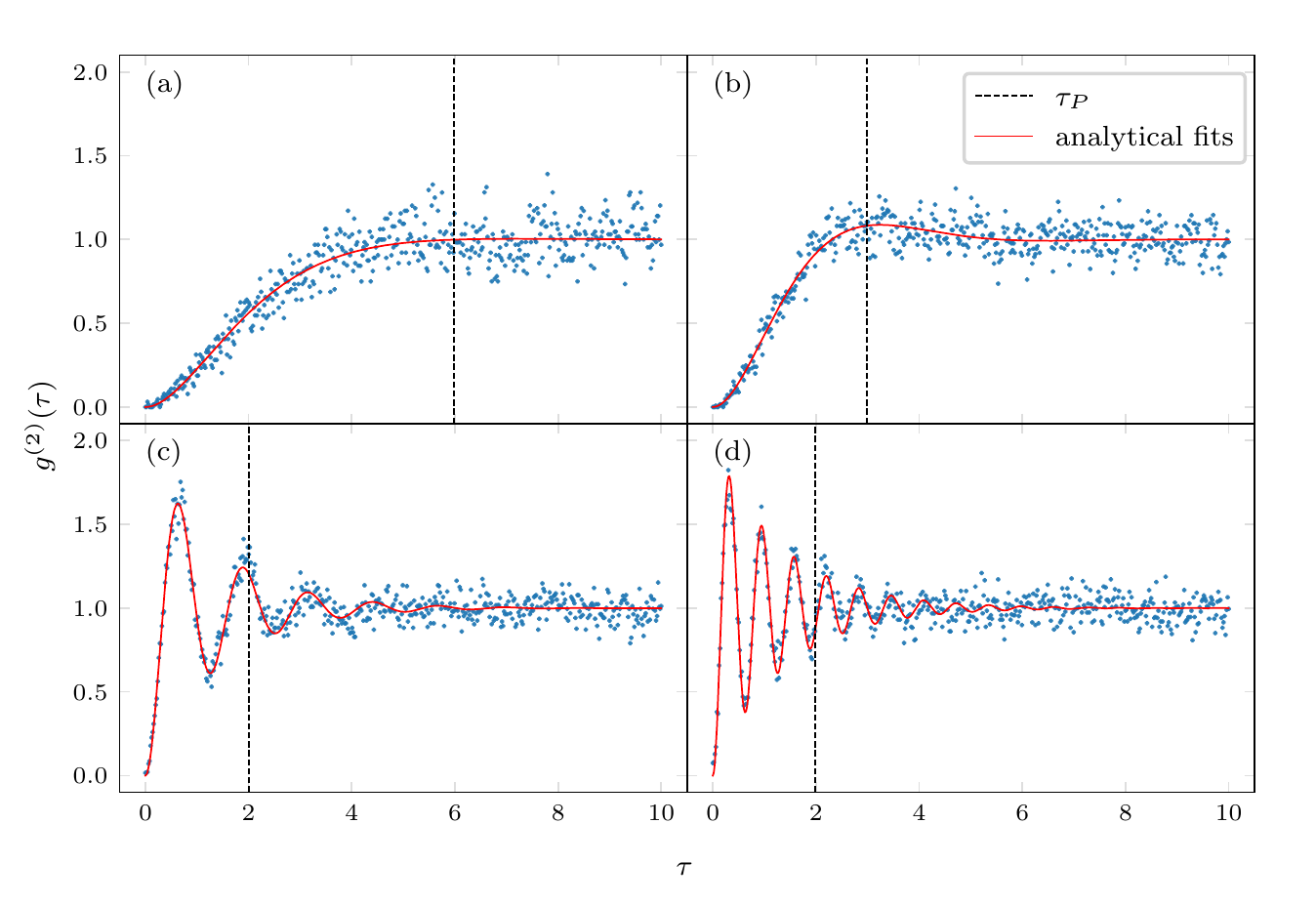}
    \vspace{-0.8cm}
    \caption{Second order coherence functions for resonance fluorescence. $\Gamma =1$ (a) $\Omega = 0.5 \Gamma$ (b) $\Omega = \Gamma$ (c) $\Omega = 5 \Gamma$ (d) $\Omega = 10 \Gamma$. Photon statistics calculated with detector fidelity $\Delta t_{D} = 0.1 \tau_{P}$, where $\tau_{P}$ is the average photon spacing, and simulation run for time $T=50,000\Gamma$.}
    \label{fig:g2s}
\end{figure}

One can obtain an analytical form for $g^{(2)}(\tau)$ by using the quantum regression theorem under the Markov approximation, %to calculate the two-time correlation function of the dipole raising and lowering operators, then, subbing the result and the equation for the electric field into Eq.(\ref{eq:quantumg2}). 
yielding the following form for the second-order coherence function for resonance fluorescence \cite{Scully},
\begin{equation}
    g^{(2)}(\tau) = 1-\text{exp} \left( -\frac{3\Gamma}{4} \tau \right) \left[ \text{cos}(\mu \tau) + \frac{3\Gamma}{4\mu} \text{sin}(\mu \tau) \right], \; \; \mu = \sqrt{\Omega^{2} - \frac{\Gamma^{2}}{16}}.
    \label{eq:g2scully}
\end{equation}
Antibunching is immediately apparent by setting $\tau$ to zero, and in the strong driving limit of $\Omega \gg \Gamma$, we  recover the cosine Rabi oscillations.
Alternatively, for the case of a single driven atom, these two features in the coherence function could have been readily predicted by simulating state evolutions, as  $g^{(2)}(\tau)$ is proportional to the probability that the system will be in the excited state at time $\tau$, given that it was initially in the ground state. \cite{GK} One can see clearly the similarities of the analytical fits in Fig.\ref{fig:g2s}, to the many-trajectory simulations of Fig.\ref{fig:DSA}. We see that the results of our computational method of calculating $g^{(2)}(\tau)$ are qualitatively in clear agreement with Eq.(\ref{eq:g2scully}), and that the quantum jump method provides a multifaceted benefit using both individual and many-trajectory simulations.

%\newpage
\section{Modelling Two-Atom Dipole-Dipole Coupled Systems}
\label{sec:Modelling2As}

For the remainder of this report, we focus on two dipole-dipole coupled two-atom systems driven by an external laser in the semi-classical regime, and consider very small interatomic separations between atoms that are assumed to be fixed both in space and orientation.
We describe each atom by the Rabi Hamiltonian of Eq.(\ref{eq:Hopt}), with the same dissipative coupling $\Gamma_{1}=\Gamma_{2}=\Gamma$ to the environment and, due to the small separation, the same coupling to the laser field $\Omega_{1}=\Omega_{2}=\Omega$. The atoms have respective dipole moments $\boldsymbol{\mu}_{1}$ and $\boldsymbol{\mu}_{2}$, and
we treat the case of distinct atomic transition frequencies of $\omega_{1}$ and $\omega_{2}$. Calculations of Lamb shifts require complex methods,\cite{Scully} so, as in Ref.\cite{Uzma}, we simply consider them as part of $\omega_{1}$ and $\omega_{2}$.
Under these assumptions, the system Hamiltonian is given by
\begin{equation}
    H_{S}=H_{1\, \text{opt}} \otimes \mathds{1}_{2} +  \mathds{1}_1 \otimes H_{2\, \text{opt}} + H_{dd},
\end{equation}
where $H_{i\,\text{opt}}$ is given by Eq.(\ref{eq:Hopt}) with $w=w_{i}$ for $i=1,2$, 
and $H_{dd}$ describes the dipole-dipole interaction between the two atoms and is given by 
\begin{equation}
    H_{dd}=\hbar \left(VS_{1}^{+}\otimes S_{2}^{-} + H.c. \right).
\end{equation}
Here, $V$ is the vacuum induced coherent dipole-dipole interaction between the two atoms, and is given by the small separation limit of the retarded dipole-dipole interaction.\cite{Varada} This becomes the static dipole-dipole interaction potential \cite{V}
\begin{equation}
    V=\frac{3\Gamma}{4(k_{0}r_{12})^{3}} [(\boldsymbol{\hat{\mu}}_{1} \cdot \boldsymbol{\hat{\mu}}_{2} ) - 3 (\boldsymbol{\hat{\mu}}_{1} \cdot \boldsymbol{\hat{r}}_{12})(\boldsymbol{\hat{\mu}}_{2} \cdot \boldsymbol{\hat{r}}_{12})].
\end{equation}
%Here, we are in a near field regime and thus $V$ scales as $1/r^{3}$. 
Thus, denoting individual ground states by $\ket{0}$ and excited states by $\ket{1}$, in the tensor product basis $\{\ket{11}, \ket{10}, \ket{01}, \ket{00}\}$, we find the system Hamiltonian is given by
\begin{equation}
    \begin{aligned}
        H_{S}=\hbar\begin{pmatrix}
        -\Delta & \Omega/2 & \Omega/2 & 0\\
        \Omega/2 & \delta/2 & V & \Omega/2 \\
        \Omega/2 & V & -\delta/2 & \Omega/2 \\
        0 & \Omega/2 & \Omega/2 & \Delta
        \end{pmatrix},
    \end{aligned}
    \quad\quad
    \begin{aligned}
        \Delta&=\omega_{L}-\frac{1}{2} \left( \omega_{1} + \omega_{2} \right), \\[1mm]
        \delta&=(\omega_{1}-\omega_{2}),
    \end{aligned}
\label{eq:Hs2Ap}
\end{equation}
where $\Delta$ is the total detuning and $\delta$ is the difference in atomic transition frequencies.

Many interesting effects arise as a result of the coherent coupling between the two atoms. Firstly, the dipole-dipole interaction induces cooperative effects that lead to greatly modified emission characteristics, and these effects are dependent on the static dipole-dipole potential $V$, present for both identical and non-identical atoms. Secondly, we find that the dipole-dipole interaction induces a coherent coupling through the vacuum field between the two intermediary states, and that the amplitude of this coupling is parameterised by the difference in atomic transition frequencies $\delta$, with $V$ playing a similar role to that of the Rabi frequency.

\subsection{Diagonalising the Lindblad master equation}
\label{subsec:DiagonalisingLindblad}
\vspace{-0.2cm}

For the two atom system, the relaxation superoperator in terms of the raising and lowering dipole operators of each atom is known to take the following form \cite{Uzma}:
\begin{equation}
\begin{aligned}
    \mathcal{L}_{\text{relax}} = &-\frac{1}{2} \Gamma \left(S_{1}^{+}S_{1}^{-}\rho_{S} +\rho_{S}S_{1}^{+}S_{1}^{-} - 2 S_{1}^{-}\rho_{S}S_{1}^{+} \right) \\[1mm]
    &-\frac{1}{2} \Gamma_{12} \left(S_{1}^{+}S_{2}^{-}\rho_{S} +\rho_{S}S_{1}^{+}S_{2}^{-} - 2 S_{2}^{-}\rho_{S}S_{1}^{+} \right) \\[1mm]
    &-\frac{1}{2} \Gamma_{12} \left(S_{2}^{+}S_{1}^{-}\rho_{S} +\rho_{S}S_{2}^{+}S_{1}^{-} - 2 S_{1}^{-}\rho_{S}S_{2}^{+} \right)\\[1mm]
    &-\frac{1}{2} \Gamma \left(S_{2}^{+}S_{2}^{-}\rho_{S} +\rho_{S}S_{2}^{+}S_{2}^{-} - 2 S_{2}^{-}\rho_{S}S_{2}^{+} \right).
\end{aligned}
\label{eq:lrelax2Ap}
\end{equation}
Here, we see the cross-terms with the $\Gamma_{12}$ cross-damping rate arising from the coupling of the bare systems through the vacuum field, where spontaneous emission from one of the atoms influences the spontaneous emission of the other.\cite{Ficek} In the limit of very small separation this cross-damping term is given by \cite{Uzma}:
\begin{equation}
    \Gamma_{12}=\Gamma(\boldsymbol{\hat{\mu}}_{1} \cdot \boldsymbol{\hat{\mu}}_{2} )
\end{equation}

It was here that we encountered the main obstacle in simulating non-identical atoms, as the quantum jump method requires the relaxation superoperator to be written in the diagonal form of Eq.(\ref{eq:mcwfrelaxationsuperoperator}). Hence, the cross-terms in Eq.(\ref{eq:lrelax2Ap}) mean that the relaxation superoperator written in the basis of each respective atom's dipole operators is not suitable for computation. 
To resolve this challenge, we note that the coefficients $\gamma_{ij}$ of Eq.(\ref{eq:relaxationsuperoperator}) can be arranged to form a Hermitian, and therefore diagonalisable, matrix $\gamma$,\cite{Manzano} and that an approach proposed by Uzma et al. in Ref.\cite{Uzma}, whereby a unitary transform of the dipole operators can be applied to obtain new symmetrised jump operators, is equivalent to diagonalising $\gamma$. As such we define new, symmetrised ``Uzma" operators of the form
\begin{equation}
    \begin{aligned}
        U_{s}^{+} &= \frac{1}{\sqrt{2}} \left( S_{1}^{+} + S_{2}^{+}  \right), \quad U_{s}^{-} = \left( U_{s}^{+}\right)^{\dagger} \\[1mm]
        U_{a}^{+} &= \frac{1}{\sqrt{2}} \left( S_{1}^{+} - S_{2}^{+} \right), \quad U_{a}^{-} = \left( U_{a}^{+}\right)^{\dagger}.
    \end{aligned}
\end{equation}
The Lindblad master equation is invariant under unitary transforms of the jump operators, and as such, we obtain the following for the relaxation superoperator
\begin{equation}
\begin{aligned}
    \mathcal{L}_{\text{relax}} = &-\frac{1}{2} \Gamma_{s} \left(U_{s}^{+}U_{s}^{-}\rho_{S} +\rho_{S}U_{s}^{+}U_{s}^{-} - 2 U_{s}^{-}\rho_{S}U_{s}^{+} \right) \\[1mm]
    &-\frac{1}{2} \Gamma_{sa} \left(U_{s}^{+}U_{a}^{-}\rho_{S} +\rho_{S}U_{s}^{+}U_{a}^{-} - 2 U_{a}^{-}\rho_{S}U_{s}^{+} \right) \to 0 \\[1mm]
    &-\frac{1}{2} \Gamma_{as} \left(U_{a}^{+}U_{s}^{-}\rho_{S} +\rho_{S}U_{a}^{+}U_{s}^{-} - 2 U_{s}^{-}\rho_{S}U_{a}^{+} \right) \to 0\\[1mm]
    &-\frac{1}{2} \Gamma_{a} \left(U_{a}^{+}U_{a}^{-}\rho_{S} +\rho_{S}U_{a}^{+}U_{a}^{-} - 2 U_{a}^{-}\rho_{S}U_{a}^{+} \right),
\end{aligned}
\label{eq:lrelax2As}
\end{equation}
where by simple algebraic comparison with the coefficients of Eq.(\ref{eq:lrelax2Ap}), we find that as both atoms have the same coupling to the environment $\Gamma$, the cross term coefficients vanish: $\Gamma_{as}=\Gamma_{sa}=0$.
Thus, we obtain the diagonalised form of the relaxation superoperator that is required for computation as
\begin{equation}
    \begin{split}
    \mathcal{L}_{\text{relax}} = -\frac{1}{2} \sum_{m=s, a} \Gamma_{m} \left(U_{m}^{+}U_{m}^{-}\rho_{S} +\rho_{S}U_{m}^{+}U_{m}^{-} - 2 U_{m}^{-}\rho_{S}U_{m}^{+} \right),
\end{split}
\end{equation}
and obtain the following symmetric and antisymmetric jump operators
\begin{equation}
\begin{aligned}
    C_{s} &= \sqrt{\Gamma_{s}}U_{s}^{-}, \quad \Gamma_{s}=\frac{1}{2} \left(\Gamma + \Gamma_{12}\right), \\[1mm]
    C_{a} &= \sqrt{\Gamma_{a}}U_{a}^{-}, \quad \Gamma_{a}=\frac{1}{2} \left(\Gamma - \Gamma_{12}\right).
    \label{eq:2Ajump}
\end{aligned}
\end{equation}
These jump operators still correspond to spontaneous emission events, and from $\Gamma_s$ and $\Gamma_a$  we see that the collective interactions between the atoms give rise not only to the coherent coupling, but also to the super- and sub-radiant modification of dissipative spontaneous emission rates.\cite{Dicke}

This diagonalisation, such that the number of coefficients is equal to the number of atoms, is a fundamental requirement of the quantum jump method of Ref.\cite{MCWF}.
However, as it was somewhat unclear when encountering this content for the first time, we think it absolutely key to emphasise for the reader that the diagonalisation above refers \textit{specifically} to the diagonalisation of the coefficients $\gamma_{ij}$. This does not correspond to changing the basis with which we represent our Hamiltonian of Eq.(\ref{eq:Hs2Ap}), or the single atom, or symmetrised jump operators $U_{m}$. Once one has written the relaxation operator with these diagonal coefficients, one is then free to choose a basis in which to represent the superoperator and system, and all accompanying physics.

\subsection{Transforming to Eigenenergy basis}
\label{subsec:TransformingtoEigenbasis}
\vspace{-0.2cm}

%Having diagonalised the relaxation superoperator, we now chose to work in the eigenenergy basis of the system Hamiltonian. 
%This allows for the relaxation superoperator to only contain terms relating to spontaneous emission, with the coherent dipole-dipole induced state-coupling fully described by the no-jump evolution. 

As we will later show, system dynamics leading to dipole-dipole induced phenomena are complex, and to demonstrate them appropriately we work in the eigenenergy basis.
From Eq.(\ref{eq:Hs2Ap}) see that the Hamiltonian is not diagonal in the absence of laser driving ($\Omega=0$), and thus that the product states are not the eigenstates of the system. As we'll now see, the two two-level-atom system is equivalent to a single, four-level system, and following a similar approach to that of Varada et al. in Ref.\cite{Varada}, we diagonalise Eq.(\ref{eq:Hs2Ap}) in the absence of laser driving to obtain the eigenstates and eigenenergies of this four-level system. Using $H_{\text{eig}}=D H_{S} D$ we obtain
\begin{equation}
    H_{\text{eig}}=\hbar\begin{pmatrix}
        -\Delta & 0 & 0 & 0\\
        0 & \lambda & 0 & 0 \\
        0 & 0 & -\lambda & 0 \\
        0 & 0 & 0 & \Delta
        \end{pmatrix}, \quad \lambda = \sqrt{\frac{\delta^{2}}{4} + V^{2}},
\end{equation}
\begin{equation}
    D=\begin{pmatrix}
        1 & 0 & 0 & 0\\
        0 & \alpha & \beta & 0 \\
        0 & \beta & -\alpha & 0 \\
        0 & 0 & 0 & 1
        \end{pmatrix}, \quad
    \alpha= \frac{1}{\left( 1 + \frac{\left(\frac{\delta}{2} - \lambda\right)^{2}}{V^{2}} \right)^{1/2}}
    ,\quad
    \beta= \frac{1}{\left( 1 + \frac{\left(\frac{\delta}{2} + \lambda\right)^{2}}{V^{2}} \right)^{1/2}},
    \label{eq:D}
\end{equation}
%and where $\alpha^{2} +\beta^{2}=1$.
This diagonalisation gives us the eigenstates and eigenenergies, shown in Fig.\ref{fig:4level}, as
\begin{equation}
\begin{aligned}
    \ket{e} &= \ket{11}   \\[1mm]
    \ket{s} &= \alpha\ket{10} + \beta\ket{01}  \\[1mm]
    \ket{a} &= \beta\ket{10} -\alpha\ket{01}  \\[1mm]
    \ket{g} &= \ket{00} 
\end{aligned}
\quad\quad\quad
\begin{aligned}
    E_{e} &= +\frac{\omega_1 + \omega_2}{2} \\
    E_{s} &= +\lambda \\[1mm]
    E_{a} &= -\lambda \\
    E_{g} &= -\frac{\omega_1 + \omega_2}{2}
\end{aligned}
\label{eq:2Aeigenstates}
\end{equation}
where the excited and ground states remain the product states of the individual atoms, and the intermediary states are given by `symmetric' and `antisymmetric' superpositions.
From this, we reintroduce the laser driving of the atom by applying the coordinate transform of Eq.(\ref{eq:D}) to Eq.(\ref{eq:Hs2Ap}) to obtain the system Hamiltonian in the eigenbasis as
\begin{equation}
    H_{S}=\hbar\begin{pmatrix}
        -\Delta & \frac{\Omega(\alpha+\beta)}{2} & \frac{\Omega(\beta-\alpha)}{2} & 0\\
        \frac{\Omega(\alpha+\beta)}{2} & \lambda & 0 & \frac{\Omega(\alpha+\beta)}{2} \\
        \frac{\Omega(\beta-\alpha)}{2} & 0 & -\lambda & \frac{\Omega(\beta-\alpha)}{2} \\
        0 & \frac{\Omega(\alpha+\beta)}{2} & \frac{\Omega(\beta-\alpha)}{2} & \Delta
        \end{pmatrix}, \quad \lambda = \sqrt{\frac{\delta^{2}}{4} + V^{2}},
\end{equation}
with this Hamiltonian being used for the remainder of this report.

\begin{figure}[b!]
    \centering
    \includegraphics[width=0.9\linewidth]{"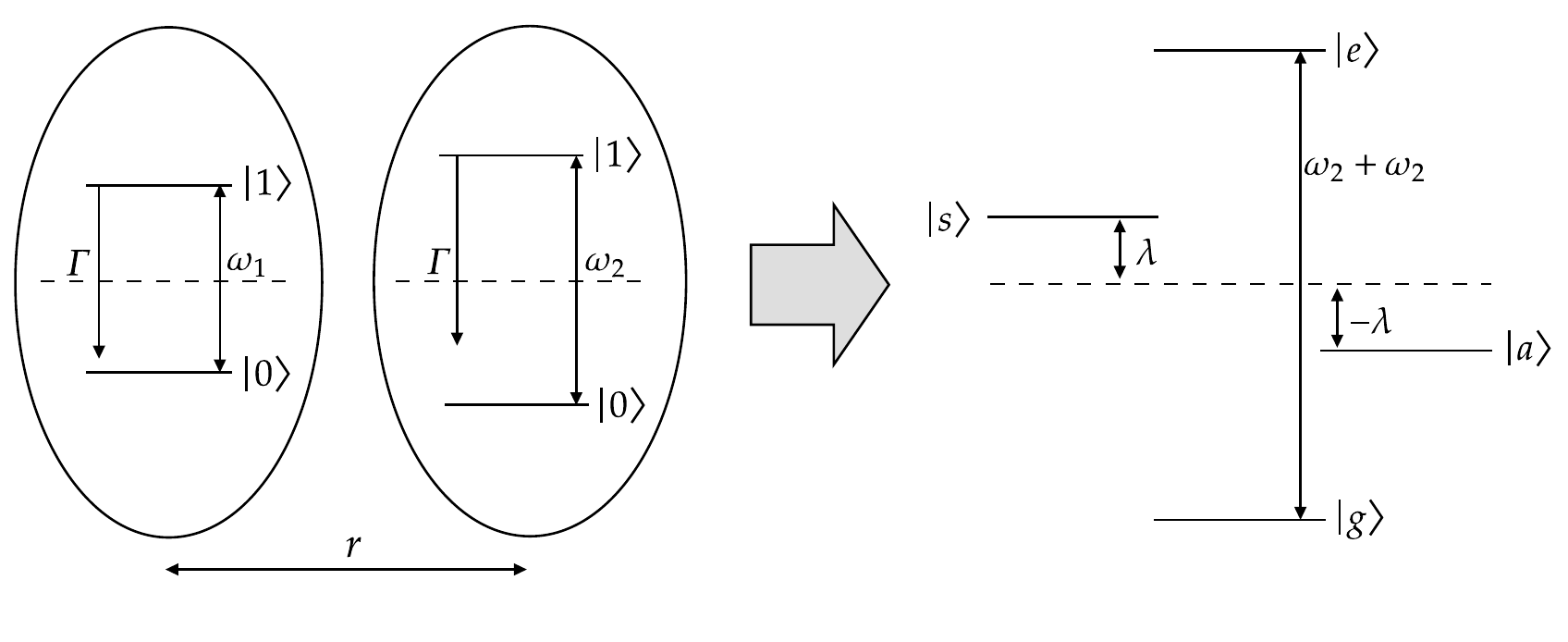"}
    \caption{Two bare two-level atoms undergoing dipole-dipole coupling to form a 4-level system. Dashed lines represents the chosen zero energy level, and the energy levels of the individual atoms are labelled numerically in increasing energy.}
    \label{fig:4level}
\end{figure}

\newpage
\subsection{How QuTip simulates}

Despite the relevant Hamiltonian for the two-atom dipole-dipole coupled system being a simple $4\times4$ matrix, the state space of the system is expansive, and as such all further simulations are now conducted using the QuTip package and `mcsolve' function to simulate the quantum jump method, as this significantly reduces the computation time required and is equivalent to the method of Mølmer et al.
The method employed by QuTip closely resembles that of those developed by Hegerfeldt and Dum \& Zoller, whereby the system wavefunction is propagated through time using the non-unitary Schrödinger equation, up to a point where the decrease in norm is equal to some random value and a quantum jump occurs.\cite{HegerfeldtQJM,Dum1,Dum2}

The algorithm to simulate a single trajectory is as follows; Step {\color{red}1.} is to choose a random number $\epsilon$ that is uniformly distributed between 0 and 1 corresponding to the probability that a quantum jump occurs. Next, Step {\color{purple}2.} is to obtain the non-unitary Schrödinger equation using the effective Hamiltonian to propagate the system wavefunction through time, and to then integrate it to some time $\tau$ when the norm squared of the wavefunction, $\braket{\psi(t) |\psi(t)}$, is equal to the random number $\epsilon$, at which point quantum jump occurs. Step {\color{blue}3.} consists of a quantum jump at time $\tau$, where a jump operator, $C_n$, to collapse to a new state is selected from the same states as Eq.(\ref{eq:quantumjumpevolution}) in the method of Mølmer et al., according to the smallest $n$ that satisfies
\begin{equation}
    \sum_{i=1}^{n} P_{n}(\tau)\geq \epsilon, \quad P_n(\tau) = \braket{\psi(\tau)|C^{\dagger}_n C_n |\psi(\tau)}/\delta p .
\end{equation}
Finally, the renormalised state from Step {\color{blue}3.} is used as the new initial condition at time $\tau$ and a new random number is generated and the above procedure repeated until the final simulation time is reached.

Alongside it's reduced computation time, the QuTip package allows for the time evolution of system states and various expectation values to be calculated, and for the generation of an array of jump times, and these tools allow us to investigate and reveal interesting emission characteristics emergent from the simple two-atom dipole-dipole coupled system.

\begin{figure}[b!]
    \centering
    \includegraphics[width=0.96\linewidth]{"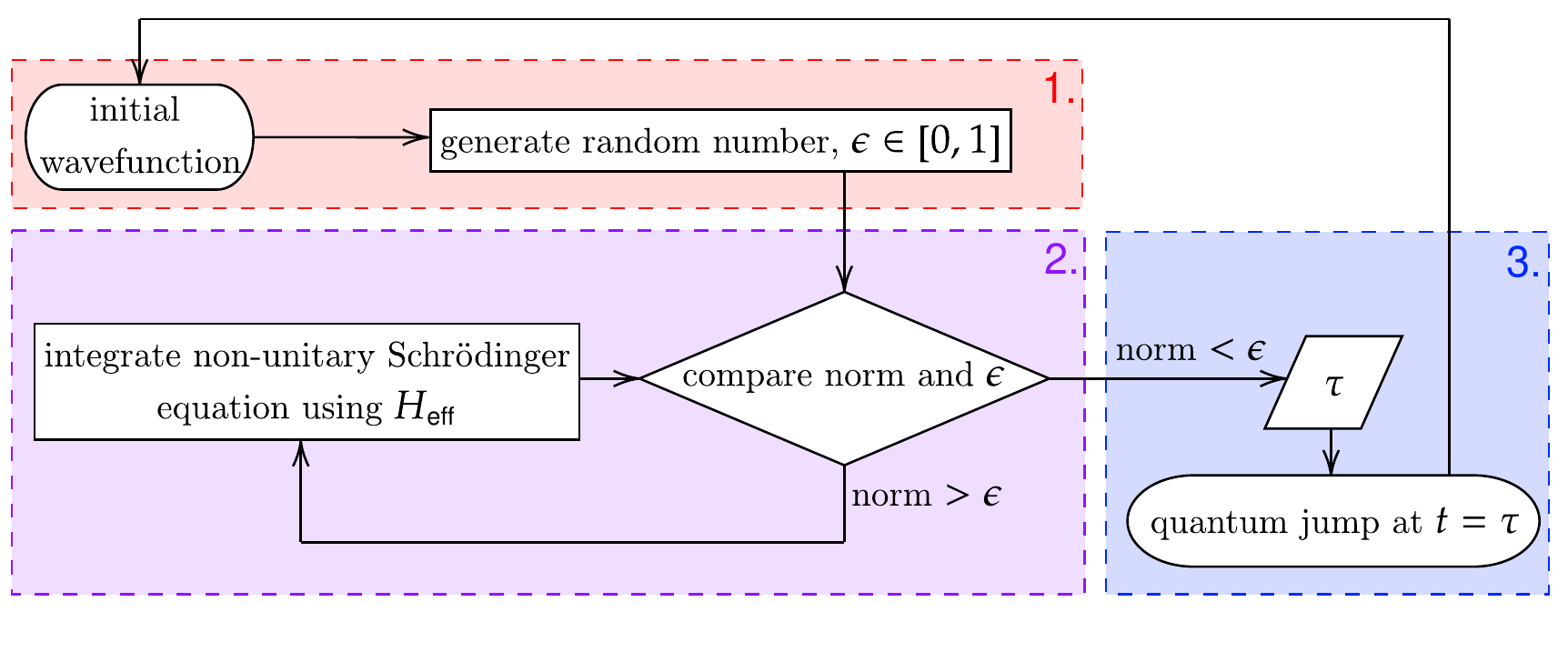"}
    \caption{Algorithm that QuTip uses to compute the quantum jump method}
    \label{fig:qutipalgorithm}
\end{figure}

\newpage

\section{Cooperative Dipole Effects} 
\label{sec:CooperativeDipoleEffects}
\vspace{-0.2cm}

In this section, we investigate cooperative effects induced by the static dipole-dipole potential $V$. In order to demonstrate the phenomena induced we replicate the results of Hettich et al. in Ref.\cite{Hettich}. To obtain the steady-state state occupations we use the QuTip steadystate($H,\{C_{m}\}$) function, and compute in both the eigenbasis as well as the product basis to provide additional insight into underlying physical phenomena.

\begin{figure}[b!]
    \centering
    \hspace{-1mm}
    \vspace{-0.3cm}
    \includegraphics[width=\linewidth]{"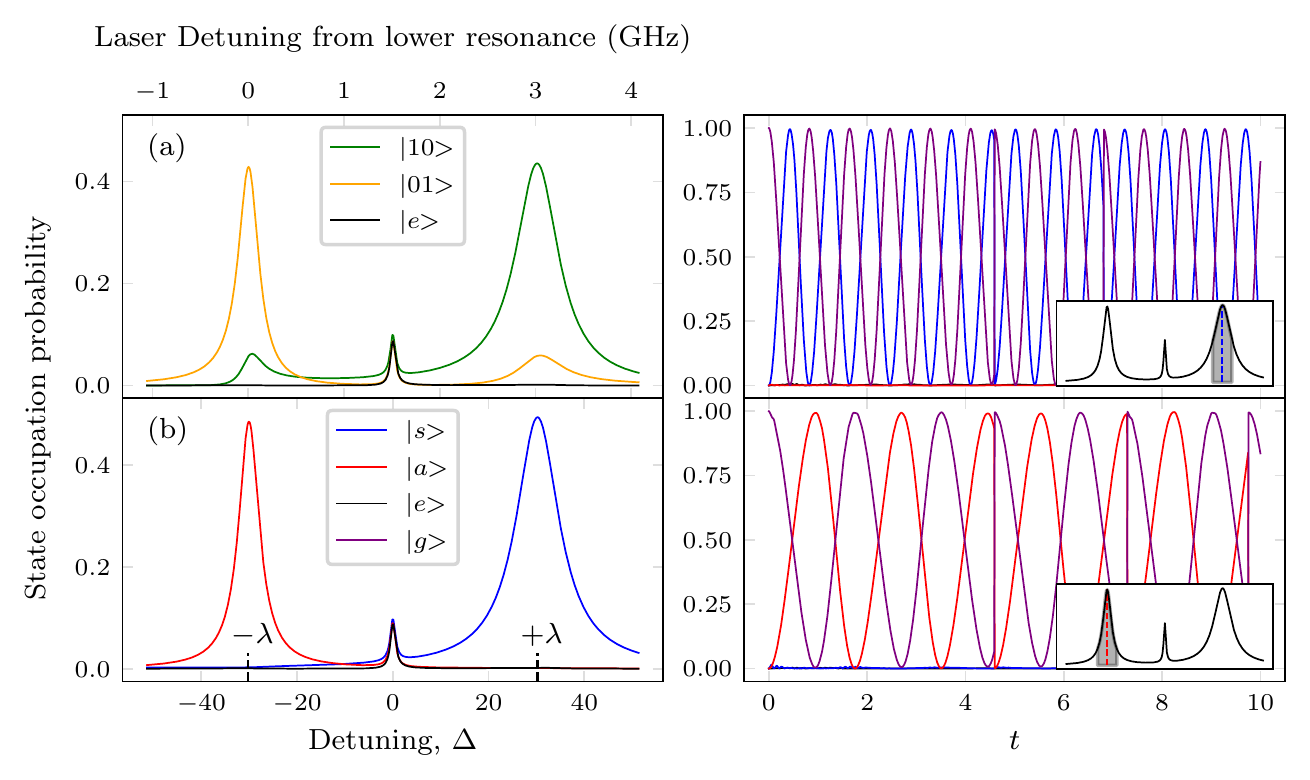"}
    \caption{\textit{Left}: Steady state occupations obtained using QuTip steadystate() in basis of (a) product state steady states as presented by Hettich et al. in GHz, (b) eigenenergy steady-states, where the ground steady state has been omitted for both. \textit{Right}: State evolutions of each intermediary resonance. Simulated using system variables rescaled by factor $\Gamma$ giving: $\Gamma=1$, $\delta=46.4$, $V=19.3$, $\Gamma_{12}=0.18$, and $\lambda=30.2$. $\Omega\approx6$ was used to obtain roughly the same profile as that of Hettich. Paramters yield $\alpha=0.94$ and $\beta=0.34$.}
    \label{fig:hettichss}
\end{figure}

When looking at the steady-state occupation of the system in Fig.\ref{fig:hettichss} we observe two larger resonances, and by transforming to the eigenbasis we see that each of these peaks occur at the transition frequencies of each of the intermediary states.
For two non-interacting atoms there would exist only these two peaks, however,  we also see the formation of a third, central resonance corresponding to when the total detuning, $\Delta$, equals zero.
The nature of these peaks is demonstrated using photon statistics in Fig.\ref{fig:hettichg2s}. In Fig.\ref{fig:hettichg2s}{\color{blue}(a)} and {\color{blue}(b)}, we see that the photon statistics of each intermediary state resonance exhibit antibunching with Rabi oscillations. Physically, as evidenced by the trajectories in Fig.\ref{fig:hettichss}, we conclude that the outer resonances correspond to the single-photon resonance fluorescence of each respective intermediary state, each lightly perturbed by the far detuned resonance of the other. Indeed, we find that the photon statistics can be reasonably modelled by the same equation as for the case of a single driven atom, Eq.(\ref{eq:g2scully}), with the states $\ket{s}$ and $\ket{a}$ having effective Rabi frequencies of $\Omega(\alpha+\beta)$ and $\Omega(\beta-\alpha)$ respectively. As this is a single-photon process, the same analysis as in Section \ref{sec:DrivenSingleAtom} applies.

In Fig.\ref{fig:hettichg2s}\color{blue}(c), \color{black}we see that the light emitted when driving the central resonance is in fact bunched, and this arises due to a dipole-dipole interaction-induced two-photon absorption and emission process. These multi-photon processes are automatically included in our simulation as Eq.(\ref{eq:mcwfrelaxationsuperoperator}) treats the external field and dipole interaction to all orders.\cite{Varada} Regarding two-photon absorption, this interaction opens a new, totally different channel of excitation, where each atom is off resonantly excited by absorption of a photon, and the coupling allows for a compensational energy transfer between the symmetric and antisymmetric states.\cite{2PAbsorption} The proceeding two-photon emission process can be neatly explained by excellent arguments presented by Beige \& Hegerfeldt in Ref.\cite{BeigePhoton}, whereby the level populations are greatly modified by the emission of a photon, leading to a higher probability density for the emission of a further photon.  Indeed, for the system investigated by Hettich, by calculating the probability of spontaneous emission in the steady-state, followed by the probability of a consecutive emission after having jumped from the steady-state, we find over an order of magnitude increase in probability of spontaneous emission of over an order of magnitude, with  $\delta p_{\text{con.}} /\delta p_{\text{ss}} \approx 30$.
\begin{figure}[b!]
    \centering
    \hspace{-1mm}
    \includegraphics[width=1\linewidth]{"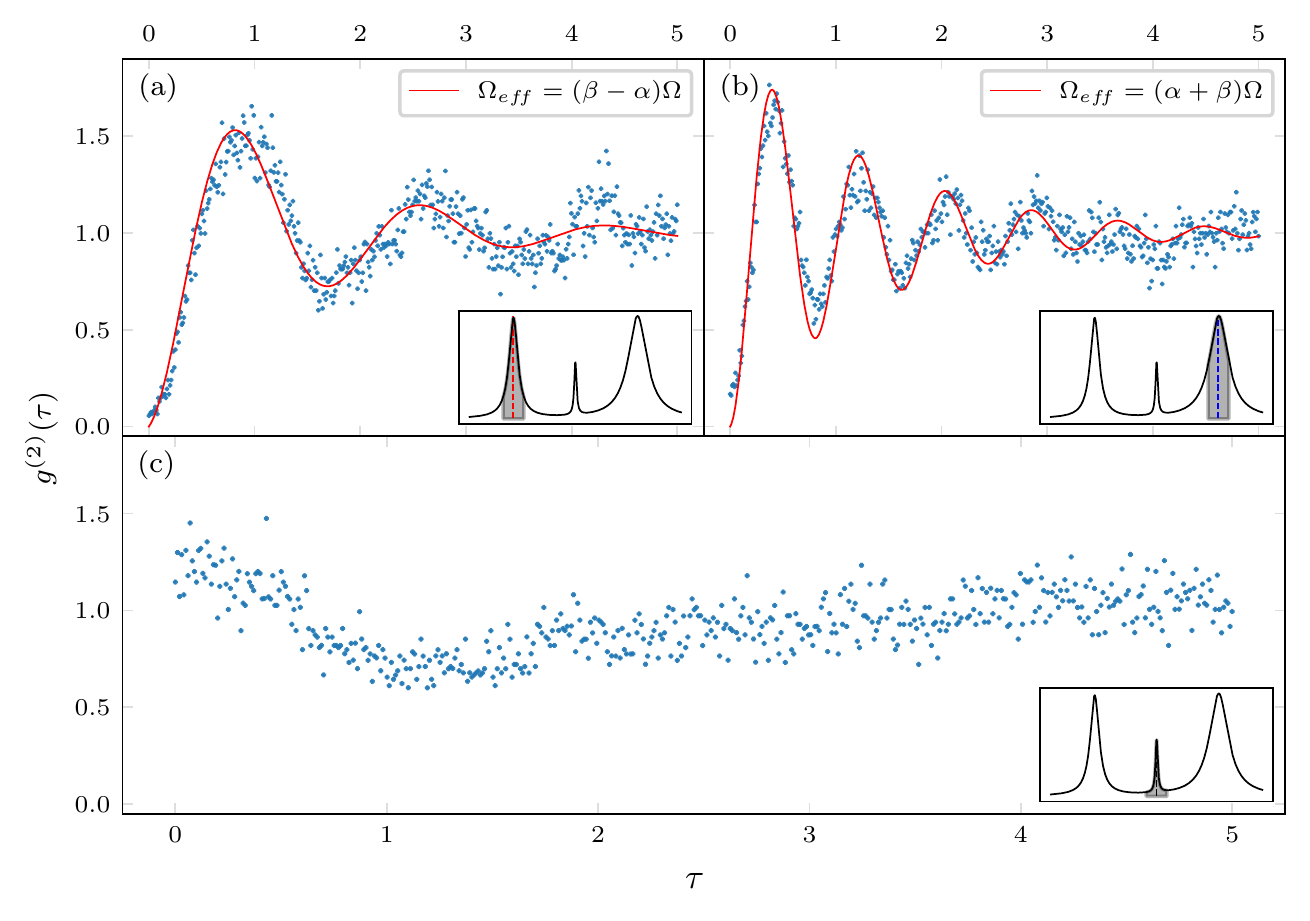"}
    \vspace{-8mm}
    \caption{Second order coherence functions for the light emitted on the three resonances in the spectrum of two-atom d-d coupled system. Resonance being scanned over inset in bottom right of each figure. (a) Antisymmetric resonance obtained at $\Delta=-\lambda$ (b) Symmetric resonance obtained at $\Delta=+\lambda$ (c) Cooperative dipole-dipole induced resonance obtained at $\Delta=0$. System variables of Fig.\ref{fig:hettichss} used. Analytical fits in (a) and (b) obtained using Eq.(\ref{eq:g2scully}).}
    \label{fig:hettichg2s}
\end{figure}
Finally, the mild dip in $g^{(2)}(\tau)$ after the bunched peak arises due to an analogous concept to ``dead time'', as a two-photon emission event depopulates the higher energy states. Unfortunately, obtaining an analytical fit for two atoms requires a remarkably high volume of algebra.\cite{Sofia}

\newpage
\section{Dark States}
\label{sec:DarkStates}

\subsection{Identical vs Non-identical Atoms}
\label{subsec:Identicalvs}
\vspace{-0.2cm}

Here, we inspect the work of Uzma et al. in Ref.\cite{Uzma} on the decoupling from interactions of the antisymmetric state.
To do this, we drive the two-atom system on the antisymmetric resonance at $\Delta = -\lambda$, and we consider again very small interatomic separations, this time with aligned dipoles such that $\Gamma_{12}=\Gamma$.
The entirety of the work of Uzma et al. on four-level two-atom systems is represented using the maximally entangled intermediary states
\begin{equation}
    \ket{s_{\text{max}}}=\frac{1}{\sqrt{2}}(\ket{10}+\ket{01}), \quad \ket{a_{\text{max}}}=\frac{1}{\sqrt{2}}(\ket{10}-\ket{01}),
\end{equation} 
corresponding to $\alpha=\beta=1/\sqrt{2}$. In this basis, $U_{s}^{+}$ and $U_{a}^{+}$ are of the following form
\begin{equation}
    U_{s}^{+} =\ket{e}\!\bra{s_{\text{max}}} + \ket{s_{\text{max}}}\!\bra{g}, \quad     U_{a}^{+} =-\ket{e}\!\bra{a_{\text{max}}} + \ket{a_{\text{max}}}\!\bra{g}.
\end{equation}
Thus, we see that the two jump operators yield two independent decay channels: $\ket{e}\to\ket{s_{\text{max}}}\to\ket{g}$ and $\ket{e}\to\ket{a_{\text{max}}}\to\ket{g}$. Furthermore, as $\Gamma_{12}=\Gamma$ we see that $\Gamma_{a}$  equals zero, and hence, the antisymmetric jump operator and the antisymmetric decay channel vanish. Thus, the maximally entangled antisymmetric state completely decouples from dissipative interactions. 

Unfortunately, from Eq.(\ref{eq:D}), we see that these maximally entangled states only correspond to the eigenstates of the system for atoms with identical atomic transition frequencies, giving $\{\ket{s_{\text{max}}},\ket{a_{\text{max}}} \}= \{\ket{s},\ket{a} \}$. For this case, we find that this antisymmetric state completely decouples from coherent interactions allowing transfer between the intermediary states, as well as from the dissipative interactions with the environment. From the viewpoint of quantum computation, whilst the decoupling from dissipative processes is useful in shielding the state from decoherence, the state should still be accessible by coherent processes.\cite{Ficek}  Thus, we extend the above treatment so as to allow us to investigate the more general system of non-identical atoms in terms of its eigenstates.
%Uzma treats non-identical system using eigenstates basis of identical system.

%\subsubsection{Non-identical}
Using our work in section \ref{sec:Modelling2As}, for the general two-atom dipole-dipole coupled system, we find that the jump operators in the eigenbasis take the following form,
\begin{equation}
    \begin{aligned}
     C_{s} &=\sqrt{\Gamma_s} \big[(\alpha+\beta)\{\ket{s}\!\bra{e}+\ket{g}\!\bra{s}\} + (\beta-\alpha)\{\ket{a}\!\bra{e}+\ket{g}\!\bra{a}\} \big] \\[1mm]
     C_{a} &=\sqrt{\Gamma_a} \big[(\beta-\alpha)\{\ket{s}\!\bra{e}-\ket{g}\!\bra{s}\} - (\alpha +\beta)\{\ket{a}\!\bra{e}-\ket{g}\!\bra{a}\} \big] \to 0,
    \end{aligned}
\end{equation}
with $\alpha$, $\beta$ given by Eq.(\ref{eq:D}), and the antisymmetric jump operator is again decoupled. 
We see that there no longer exist independent symmetric/antisymmetric decay channels. Thus, both transitions from upper to intermediate, and from intermediate to lower states are correlated, meaning that it is impossible for us to fully decouple the antisymmetric state from interactions for the case of non-identical atoms.

As stated when modelling the two-atom system, further to reintroducing the antisymmetric excitation channel in the second half of $C_{s}$, we find that the consequence of the different atomic transition frequencies is to reintroduce the coherent population transfer between the two intermediary states.
This is demonstrated by Uzma et al. by calculating the steady-state populations, as seen in Fig.\ref{fig:zerodelta}{\color{blue}(a)}. Here, we find a greatly suppressed, however non-zero, steady-state symmetric population about the antisymmetric resonance in Fig.\ref{fig:zerodelta}{\color{blue}(a)}. In Ref.\cite{Uzma}, this suppression is interpreted as complete population transfer from $\ket{s}$ to $\ket{a}$, however, the actual system evolution is much more nuanced, and the steady-state profile which describes an ensemble average does not paint the full picture. 

As can be seen in the right-hand side trajectories of Fig.\ref{fig:UzmaEvo}, by simulating an individual trajectory we find that, unlike the steady-state profile of  Fig.\ref{fig:zerodelta}{\color{blue}(a)}, the symmetric suppression is not complete, and that there is a non-zero symmetric population undergoing complex state dynamics. We see that the system undergoes no jump evolution determined by the effective Hamiltonian, interrupted at random points by spontaneous emission events, resetting this no jump evolution. Under this no jump evolution, we see a short period of initial, off-resonant oscillation between symmetric and ground states, which becomes damped and dominated by longer resonant oscillations. % induced by the coupling of the two intermediary states. 
We believe either that the far off-resonant laser driving between the ground, symmetric and excited states leads to gradual accumulation of a laser-decoupled antisymmetric state. Or, that both the antisymmetric and symmetric states couple weakly to the laser, with the coherent transfer between the two damping, then dominating the symmetric state into out of phase oscillations with the antisymmetric state. Unfortunately, as of writing, we have been unable to obtain closed forms for the evolved wavefunction to determine the exact nature of these dynamics, for reasons discussed in Section \ref{sec:FurtherWork}.

Besides the complex dynamics revealed, the absence of any coherent transfer process for identical atoms is evident in both the lower steady-state profile and single trajectory of Fig.\ref{fig:zerodelta}. Furthermore, it is clear in both Fig.\ref{fig:zerodelta}{\color{blue}(a)} and {\color{blue}(b)} that the central resonance arising from cooperative effects remains for both non-identical and identical atoms.

\begin{figure}[b!]
    \centering
    \includegraphics[width=\linewidth]{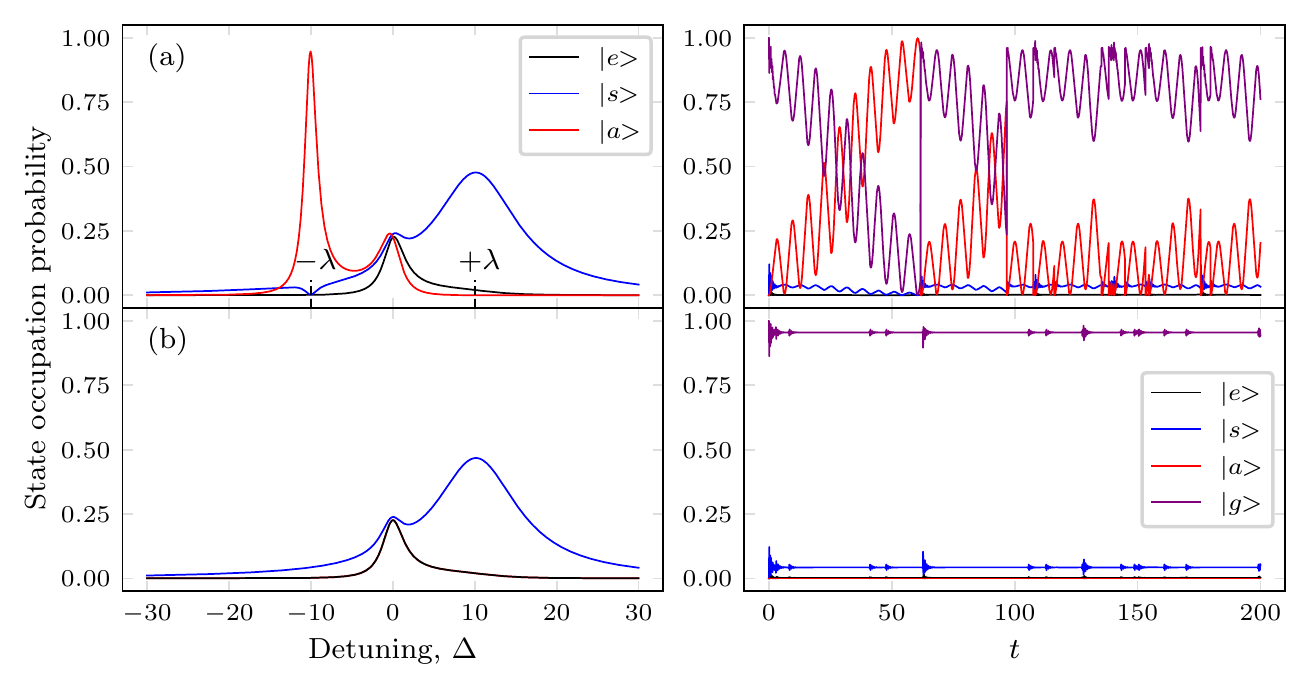}
    \vspace{-0.5cm}
    \caption{Steady state profiles and state evolutions of (a) non-identical atoms, and (b) identical atoms. Single trajectories obtained by simulating at $\Delta=-\lambda$, with system variables rescaled by factor $\Gamma$ giving: $\Gamma=1$, $V=10$, $\Gamma_{12}=1$, $\Omega=5$, and (a) $\delta=2$, (b) $\delta=0$.}
    \label{fig:zerodelta}
\end{figure}

\newpage

\subsection{Macroscopic Dark Periods}
\label{subsec:MacroscopicDarkPeriods}
\vspace{-0.2cm}

As we will now see, the origin of the significant steady-state symmetric suppression about the antisymmetric resonance can be found in the extended time evolution of the system. 
For non-identical atoms, we saw that it is possible to transfer population to the antisymmetric state both by spontaneous emission from the excited state, as occurs during a quantum jump, and by non-dissipative interactions, as occurs in the no jump evolution. Thus, whilst we do not have true dark states, as the antisymmetric state is accessible to dissipative processes as well as coherent processes, in Fig.\ref{fig:UzmaEvo} the quantum jump method reveals a complex picture of macroscopic dark states that exist over geological timescales, interspersed with light periods of complex state dynamics.

Working again in the same system configuration of Uzma et al., in Fig.\ref{fig:UzmaEvo} we see that the non-identical dynamics lead to the gradual accumulation of population in the antisymmetric state until the system enters a macroscopic dark state with no emissions. This dark state is equal to the eigenstate of $H_{\text{eff}}$ whose eigenvalue carries the smallest imaginary part, and these values obtained from $H_{\text{eff}}$ are annotated on the upper-right trajectory of Fig.\ref{fig:UzmaEvo}.

\begin{figure}[b!]
    \centering
    \includegraphics[width=\linewidth]{"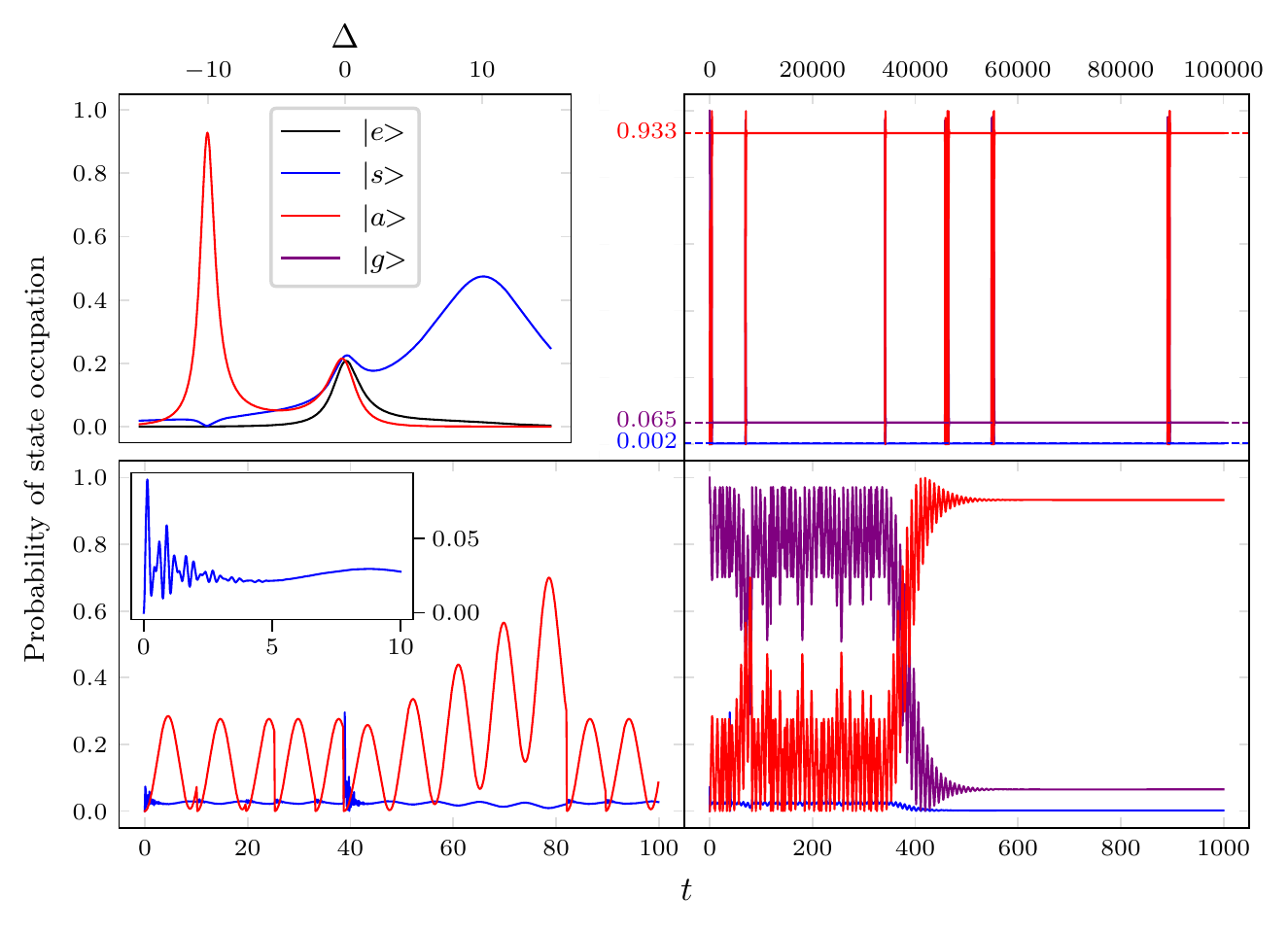"}
    \vspace{-1cm}
    \caption{\textit{Upper-left}: Steady-state occupations of system with ground state omitted.\newline \textit{Else}: State evolutions of system driven on antisymmetric resonance for various time scales. Simulated using system variables rescaled by factor $\Gamma$ giving: $\Gamma=1$,  $V=10$, $\Gamma_{12}=1$, $\Omega=5$, and $\delta=2$. Dark state predicted by $H_{\text{eff}}$ annotated on upper-right plot. }
    \label{fig:UzmaEvo}
\end{figure}

\subsubsection{No photon probability}
To begin to quantify and understand these macroscopic dark periods, given the complexity of the state dynamics, we adopt a statistical approach by using the probability $P_{0}(t;\ket{\psi_0})$ that no photon is emitted until a time $t$. This is given by the norm-squared of the system wavefunction at time $t$, which is obtained by applying the non-unitary time operator describing the evolution of an initial state wavefunction under the effective Hamiltonian, to give 
\begin{equation}
    P_{0}(t;\ket{\psi_0})= \left\lvert\left\lvert \psi (t)\right\rvert\right\rvert^{2}=\left\lvert\left\lvert\text{exp}\{-i H_{\text{eff}}t /\hbar\ket{\psi_0} \} \right\rvert\right\rvert^{2},
    \label{eq:P0}
\end{equation}
where $\ket{\psi_0}$ is some initial state at $t=0$.\cite{HegerfeldtQJM,BeigePhoton}
As proposed by Hegerfeldt in Ref.\cite{HegerfeldtQJM}, by now randomly sampling values of $P_0$ to obtain corresponding jump times and evolving the system wavefunction to these times under $H_{\text{eff}}$ we obtain a method of simulating quantum trajectories, and with $P_0$'s equivalence to the time evolution of the norm-sqaured of the system wavefunction, by treating $\ket{\psi_0}$ as the reset state after a quantum jump we see that this method is directly equivalent to that used by QuTip. As such, we are able to use $P_0$ to provide insight into and allow us to predict and quantify the formation of macroscopic dark periods. To do so for the systems considered, we seek to simplify Eq.(\ref{eq:P0}) by first noting that as the effective Hamiltonian is non-Hermitian the right eigenvectors are not pairwise-orthogonal, and 
therefore in order to diagonalise the exponential operator we introduce the dual basis states that are the left eigenvectors of the effective Hamiltonian. This allows us to rewrite Eq.(\ref{eq:P0}) as
\begin{equation}
    P_{0}(t;\ket{\psi_0})=\left\lvert\left\lvert\sum_{m}e^{-i \lambda_{m} t} \braket{\lambda^m|\psi_0} \ket{\lambda_m}\right\rvert\right\rvert^{2} =\sum_{m}\sum_{n}e^{i \lambda_m^* t}e^{-i \lambda_n t} \braket{\psi_0|\lambda^m} \braket{\lambda_m|\lambda_n} \braket{\lambda^n|\psi_0} , 
    \label{eq:P0sumlong}
\end{equation}
where $\lambda^m$ and $\lambda_m$ denote the left and right normalised eigenvectors of $H_{\text{eff}}$, with respective eigenvalues denoted by $\lambda_{m}$.
%By writing Eq.\ref{eq:P0sum} as a sum over the indices $m,n$,
%\begin{equation}
 %   P_{0}(t;\ket{\psi_0})=\sum_{m}\sum_{n}e^{i \lambda_m^* t}e^{-i \lambda_n t} \braket{\psi_0|\lambda^m} \braket{\lambda_m|\lambda_n} \braket{\lambda^n|\psi_0},
%\end{equation}
This cannot in general be further simplified as $\braket{\lambda_m|\lambda_n} \neq \delta_{m,n}$, however, in the region of the parameter space that we simulate within, the right eigenvectors are in fact to a good approximation pairwaise orthogonal.[\ref{app:ortho}]
Combined with the fact that that all $\lambda_{m}$ have negative imaginary parts,\cite{NegativeIm} we are able to a good approximation simplify $P_0$ to
\begin{equation}
\begin{aligned}
    P_{0}(t;\ket{\psi_0})=\sum_{m}e^{-2\left\lvert \mathcal{I}m(\lambda_i)\right\rvert t} \left\lvert \braket{\lambda^{m}|\psi_0} \right\rvert^{2}.
    \label{eq:P0sum}
\end{aligned}
\end{equation}
%\subsubsection{link to qutip, intuitive picture, two timescales}

%The formation of macroscopic light and dark periods can be seen directly in the no photon probability distribution by noting that due to $P_{0}$'s correspondence to the non-unitary time evolution of the initial state wavefunction under $H_{\text{eff}}$, randomly sampling according to $P_0$ to obtain corresponding jump times, and evolving the system wavefunction to these times under $H_{\text{eff}}$, is directly equivalent to the method which QuTip uses to simulate the quantum jump method.
%ime evolution of the system is equivalent to randomly sampling times according to the probability distribution P0?¿ 

\begin{figure}[t!]
    \centering
    \includegraphics[width=\linewidth]{"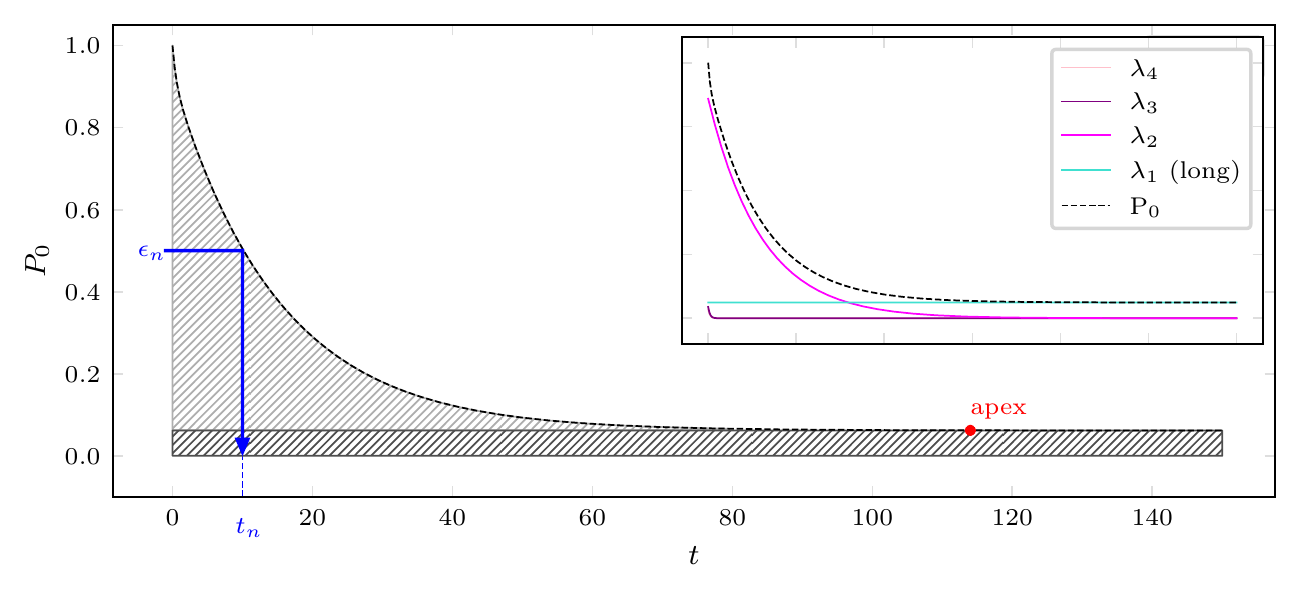"}
    \caption{The no photon probability distribution for the system configuration of Fig.\ref{fig:UzmaEvo}. Randomly sampled number $\epsilon_n$ and corresponding jump time $t_n$ shown in blue, with light and dark regions of sampling $P_0$ shown in light and dark grey respectively. Inset upper-right is the decomposition of the no photon probability distribution using Eq.(\ref{eq:P0sum}), showing the presence of the two, largely separated timescales. We see $P_0$ largely dominated by the longest lived exponentials with eigenvalues $\lambda_1$, $\lambda_2$, with a long tail formed by the longest exponential.}
    \label{fig:P0sampling}
\end{figure}

By numerically obtaining the eigenvectors and eigenvalues of $H_{\text{eff}}$ to obtain $P_0$ for the case of our driven dipole-dipole coupled two-atom system,  we find that macroscopic light and dark periods form as a result of the presence of a long tail and kink in the probability distribution when obtaining jump times. From Eq.(\ref{eq:P0sum}) we see that this tail and kink arise as a result of two largely-separated timescales in the probability distribution, seeing in the upper-right inset of Fig.\ref{fig:P0sampling} that there exists one timescale that is very slowly decaying in comparison to the other components in the summation of $P_0$. 
In order to distinguish between the two separated timescales we identify an apex in $P_0$, and the process leading to dark period formation, illustrated in Fig.\ref{fig:P0sampling}, can then be explained as follows; when sampling the distribution to obtain jump times, if our randomly generated number used to sample the distribution falls within the light section of the possible values of $P_0$ that are above the apex of the kink, then we obtain samples from a rapidly decaying exponential distribution with short times between photon emissions. However, if a random number falls within the dark section below the apex of the kink, it yields a greatly extended value of $t$ for the time between next-photon emission, hence giving rise to macroscopic dark periods in the fluorescence of the system. This picture of sampling intuitively explains the macroscopic quantum jumps between fluorescence and dark states observed in the system, and gives us a simple criteria with which to define these light and dark periods; if the first photon after state reset is emitted after the value of $t$ at the apex of the kink, $T_{apex}$, we call it a dark period, and successive photons less than $T_{apex}$ apart are denoted as a light period.

The evolution to the dark state seen in the system dynamics of Fig.\ref{fig:UzmaEvo} can likewise by readily explained by $P_0$'s correspondence to norm of the system wavefunction at time $t$ by using the middle equality in Eq.(\ref{eq:P0sumlong}). Here, we see that if there is a sufficiently long timescale as to form a long tail, then for large $t$ the exponential factors of all but the longest timescale components in the summation will tend to zero, leaving solely this longest lived element. As the state dynamics are normalised at each timestep, the prefactors in the summation are replaced by unity and the resulting system state tends towards the right eigenvector whose eigenvalue carries the smallest imaginary part.

\newpage
\subsection{Waiting Time Distribution}

Whilst the no photon probability distribution provides an intuitive, qualitative picture of light and dark period formation, it can also be used to obtain a quantitative description of their formation, and to calculate the average lengths of both of these periods by obtaining the so called waiting time distribution. This waiting time distribution gives us the probability density of spontaneous emission as a function of time, and is given by the negative of the time derivative of $P_0$ as
\begin{equation}
    w_1(t) = -\frac{d}{dt}P_{0}(t;\ket{\psi_0})= \sum_{m}\,2\left\lvert \mathcal{I}m(\lambda_m)\right\rvert e^{-2\left\lvert \mathcal{I}m(\lambda_m)\right\rvert t}\left\lvert \braket{\lambda^m|\psi_0} \right\rvert^{2} , \quad m=1,2,3,4,
    \label{eq:waitingtimedist}
\end{equation}
where we see that the distribution is the sum of four exponential distributions, weighted according to the overlaps of their respective left-eigenvectors with the reset state.

In this context of the waiting time distribution, to again understand the formation of dark periods we follow Ref.\cite{DarkPeriods} and start by denoting the eigenvalue of $H_{\text{eff}}$ with the smallest imaginary part (and its associated eigenstate) by index $m=1$. As all $\lambda_{m}$ have negative imaginary parts,\cite{NegativeIm} if $\text{Im}(\lambda_{1})$ is very small compared to that of the other eigenvalues, and $\braket{\lambda^1|\psi}\neq 0$, then there is the possibility of extended dark periods as we have a non-zero probability of exceeding the so called waiting time, $T_{apex}$, and reaching the region of the long tail where $P_{0}$ is dominated by the smallest eigenvalue and changes very little in time. This yields very small values of $w_1$ and consequently we enter a dark state with very small probability of state reset.

\begin{figure}[b!]
    \centering
    \includegraphics[width=\linewidth]{"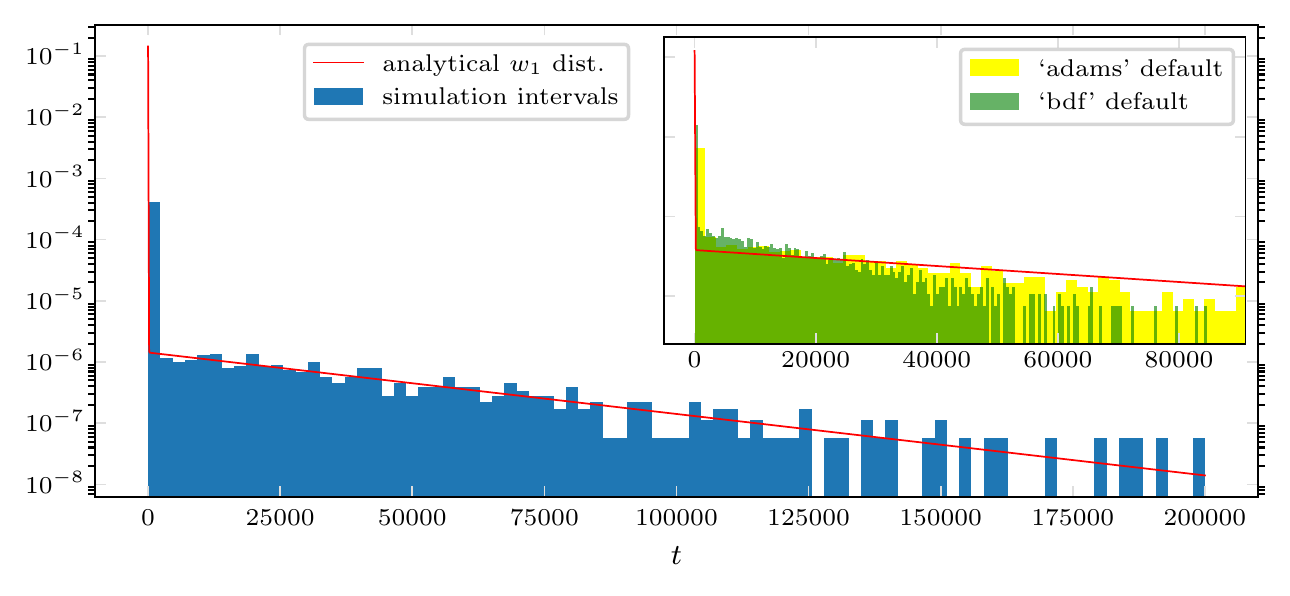"}
    \caption{Histogram of simulation intervals between photons emissions plotted in blue against the analytical waiting time distribution in red. Simulated using system variables rescaled by factor $\Gamma$ giving: $\Gamma=1$,  $V=10$, $\Gamma_{12}=1$, $\Omega=5$, and $\delta=2$, and simulation run for $t=20,000,000$ with mcsolve's ODE  solver Options arguments `rtol'=1e-10, `atol'=1e-10 to give $N=7628$ emissions and $\text{int}(\sqrt{N})=87$ histogram bins. \textit{Inset upper-right}: Errors in waiting time distributions generated using default `rtol' and `atol' parameters.}
    \label{fig:waitingdist}
\end{figure}

However, here a discrepancy is found between our simulations and the analytical predicted waiting time. By plotting a normalised histogram of the waiting times between our simulated emission times  the characteristic two-timescales of the system are observed, however, inset in Fig.\ref{fig:waitingdist} we see that the lifetimes of the long tails in the distributions generated from simulations are significantly shorter lived. This occurred for both of the methods provided for the ODE solver, ``adams'' and ``bdf'', with the former being the default solver, and the later ``bdf'' solver exacerbating this underestimation of the longest timescale, leading us to conclude that the disparity between simulation and analytical fit is due to the numerics of the default integration method used by QuTip, caused by the existence of such widely distinct timescales in the effective Hamiltonian.
This discrepency was solved by changing the relative tolerance and absolute tolerance arguments in the options for the mcsolve ODE solver, `rtol' and `atol', from 1e-6 and 1e-8, to 1e-10 and 1e-10, such that these tolerances are finer than the ratio of the smallest to longest timescale.

\subsubsection{light and dark distributions}

In order to obtain expressions for the average lengths of both light and dark periods from the waiting time distribution we split the distribution at the apex, at time $T_{apex}$, into two separate ``light" and ``dark" distributions.
To obtain the position of this apex, we look at the four components of the sum in $P_0$ and note that in our parameter space the two shortest exponentials, $m=3, 4$, decay rapidly and have very small weightings. Thus, using solely the two longest exponentials we define a value of $T_{apex}$ as the first point at which the weighted longest lived exponential is an order of magnitude larger than the next longest,
%Initial ratio is given by
%\begin{equation}
 %   \frac{\left\lvert \braket{\lambda^2|\psi}\right\rvert^{2}}{\left\lvert \braket{\lambda^1|\psi}\right\rvert^{2}}
%\end{equation}
%want ratio of expoentials to be the inverse of this at tapex:
%\begin{equation}
 %   \frac{e^{-2 \left\lvert \mathcal{I}m(\lambda_1) \right\rvert T_{apex}}\left\lvert \braket{\lambda^1|\psi}\right\rvert^{2} }{e^{-2 \left\lvert \mathcal{I}m(\lambda_2) \right\rvert T_{apex}}\left\lvert \braket{\lambda^2|\psi}\right\rvert^{2}}=\frac{\left\lvert \braket{\lambda^2|\psi}\right\rvert^{2}}{\left\lvert \braket{\lambda^1|\psi}\right\rvert^{2}}
%\end{equation}
giving
\begin{equation}
    T_{apex}=\text{log}\left(10\frac{\left\lvert \braket{\lambda^2|\psi}\right\rvert^{2} }{\left\lvert \braket{\lambda^1|\psi}\right\rvert^{2}} \right)/ 2 \left[ \left\lvert \mathcal{I}m(\lambda_2) \right\rvert - \left\lvert \mathcal{I}m(\lambda_1) \right\rvert\right] .
    \label{eq:Tapex}
\end{equation}
As demonstrated in \ref{app:Tapex}, there is in fact significant scope for choosing $T_{apex}$ whereby the the calculated average periods are the same, and as such a large range of acceptable formulae for the position of the apex.

From $T_{apex}$, we obtain the value of $P_0$ at the apex, which, from the random sampling of $P_0$'s equivalence to simulating the quantum jump method, gives us the probability of a dark period occurring after a spontaneous emission event. Using this we obtain the average dark period length by truncating the waiting time distribution and renormalising it using $P_0(T_{apex})$ to give
\begin{equation}
    T_{D}= \int_{T_{0}}^{\infty} dt \, t\, w_{1}(t)/P_{0}(T_{apex}),
\end{equation}
and with Eq.\ref{eq:waitingtimedist} we obtain
\begin{equation}
    T_{D} = \frac{1}{P_0 (T_{apex})}\sum_{m} \left\lvert \braket{\lambda^m|\psi} \right\rvert^{2} \left[T_{apex} + \frac{1}{2\left\lvert \mathcal{I}m(\lambda_i) \right\rvert }\right] e^{-2\,T_{apex}\left\lvert \mathcal{I}m(\lambda_i) \right\rvert}.
    \label{eq:TD}
\end{equation}
%For two largely seperated timescales this can be simplfied further by noting that for a sufficiently slowly decaying exponential, P0apex=u1, and that as all shorter lived timescales have decayed to many orders of magnitude smaller than the longest the sum is dominated solely by m=1. This simplifies the above equation to 
%\begin{equation}
 %   T_{D}\approx T_{apex} + \frac{1}{2\left\lvert \mathcal{I}m(\lambda_1) \right\rvert }.
%\end{equation}

For a light period where successive photons are less than $T_{apex}$ apart we obtain the mean time between emissions in the light period by similarly taking the portion of the waiting time distribution before $T_{apex}$ and rescaling. From this we obtain the average time between photons in a light period, $\tau_{L}$, as
\begin{equation}
    \tau_{L}= \int_{0}^{T_{0}} dt \, t\, w_{1}(t)/[1-P_{0}(T_{apex})],
\end{equation}
and with Eq.\ref{eq:waitingtimedist} we obtain
\begin{equation}
    \tau_{L} = \frac{1}{[1-P_0 (T_{apex})]}\sum_{m} \left\lvert \braket{\lambda^m|\psi}\right\rvert^{2} \left\{ \frac{1}{2\left\lvert \mathcal{I}m(\lambda_i) \right\rvert } - \left[T_{apex} + \frac{1}{2\left\lvert \mathcal{I}m(\lambda_i) \right\rvert } \right] e^{-2\,T_{apex}\left\lvert \mathcal{I}m(\lambda_i) \right\rvert}\right\}.
    \label{eq:tauL}
\end{equation}
The average duration of a light period is then obtained by noting that in the context of uniformly sampling from $P_0$ we expect to have to pick $n_L=1/P_0(T_{apex})$ number of photons before we select a random number that falls within the dark region, giving
\begin{equation}
    T_{L}=n_{L}\times \tau_{L} = 1/P_0(T_{apex}) \times \tau_{L} .
    \label{eq:TL}
\end{equation}

As described by the above processes, the splitting of the waiting time distribution into the light and dark distributions is shown in Fig.\ref{fig:splitdist}.
In the light distribution we see lingering signs of issues with the numerics of QuTip as we see that the shortest timescales in $w_1$ are not exhibited in the photon intervals produced, and that this leads to an overestimation in the average time between photons in a light period giving $\tau_L=14.8\pm0.2$ which is in disagreement with the analytical prediction of 13.8. Despite this discrepancy, the waiting times in the dark distribution are well matched to the analytical distribution, yielding excellent agreement in average dark period length of $T_D=42000\pm2000$ with the analytical prediction of 43000.

\begin{figure}[h!]
    \centering
    \includegraphics[width=\linewidth]{"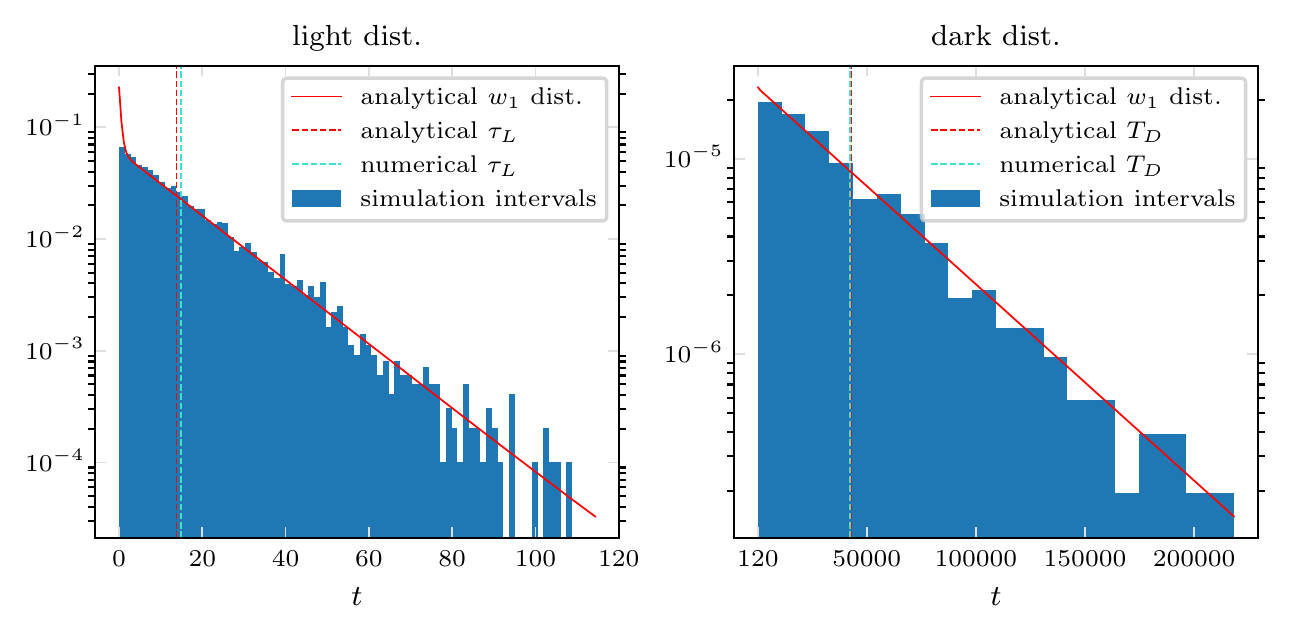"}
    \caption{Light and dark distributions obtained by splitting the waiting time distribution and simulation intervals of Fig.\ref{fig:waitingdist} at $T_{apex}\sim114$, giving 7155 photons in 84 bins in the light distribution, and 473 photons in 21 bins in the dark distribution, with histogram bin number given by $\sqrt{N}$ . Numerical averages obtained by taking the arithmetic mean of the respective photon counts. }
    \vspace{-0.5cm}
    \label{fig:splitdist}
\end{figure}

\newpage
\subsection{Macroscopic Jump Visibility}

As we have now examined the formation of these macroscopic jumps between light and dark periods of the system, we can now consider how different system configurations may effect the visibility of these jumps. This is guided by the creation of the heatmaps in Fig.\ref{fig:heatmaps} of the derived equations for period lengths.
To illustrate the the visibility of the macroscopic jumps within the emission spectra of the dipole-dipole coupled two-atom system we generate the photon streams of Fig.\ref{fig:photonstreams}.
\begin{figure}[h!]
    \centering
    \includegraphics[width=\linewidth]{"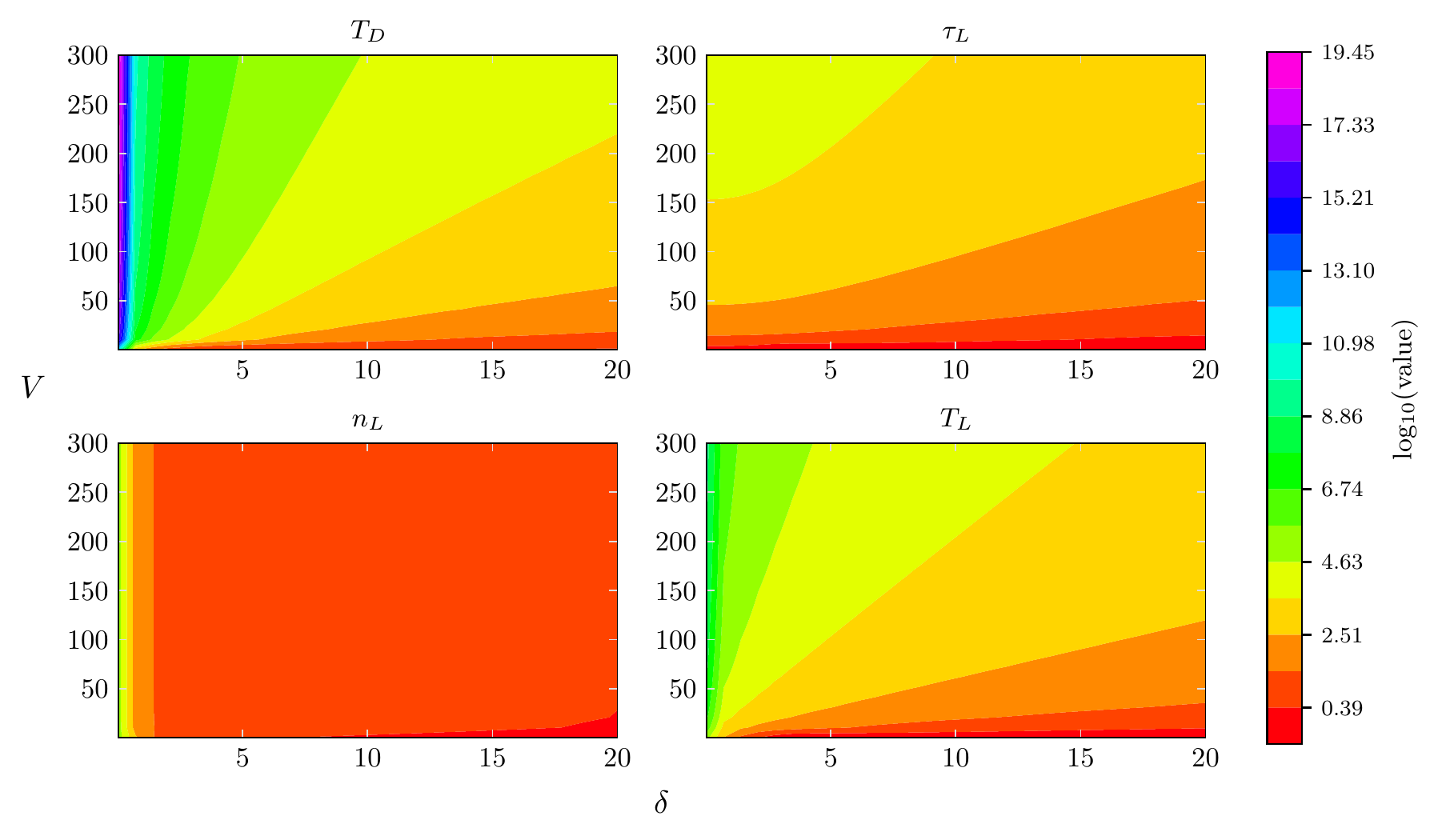"}
    \caption{Heatmaps of macroscopic parameters calculated using equations \ref{eq:Tapex}, \ref{eq:TD}, \ref{eq:tauL} \& \ref{eq:TL}. Values plotted in log$_{10}$.}
    \label{fig:heatmaps}
\end{figure}

The previous analysis for dark and light periods was contingent on small $\mathcal{I}m(\lambda_{1})$ leading to two widely different timescales, but there in fact exist three possible regimes when considering relevant magnitude of $\delta$ compared with other system parameters. 
The first is our case of the system parameters used by Uzma et al., which is an intermediary regime where $\delta$ is of the same order as other parameters. We find that this yields short light periods of photon bursts, with comparatively longer dark periods, but that the shortness of these bursts of photons is undesirable experimentally at it greatly limits visibility, as evident in Fig.\ref{fig:photonstreams}{\color{blue}(a)}.
In stream {\color{blue}(b)} we demonstrate a second regime, where $\delta$ is increased to an order of magnitude larger than other variables. We again find short bursts of photons, however this time with extremely short dark periods. Here, there is insufficient separation in the two-timescales of $P_0$, and the light and dark states merge, thus yielding no observable macroscopic jumps in the system. Fig.\ref{fig:photonstreams}{\color{blue}(c)} demonstrates the third regime, with $\delta$ an order of magnitude smaller than other parameters leading to a substantial increase in both dark and light period lengths. However, this again can lead to the same result as in {\color{blue}(b)} whereby the macroscopic jumps between light and dark periods are hard to observe, except this is now due to the extreme length of the dark periods formed of order $\sim 1\times 10^{10}$.
Whilst hard to distinguish from the photon stream figures, we also note that as predicted in the heatmap of $\tau_L$ in Fig.\ref{fig:heatmaps}, varying $\delta$ yields little change in the average time between photons in streams {\color{blue}(a)} and {\color{blue}(c)}, however, that despite the reduced number of photons per light period {\color{blue}(b)} appears very bright due to the greatly increased number of these photon bursts and the decrease in average time between their constituent photons.

As mentioned in Section \ref{sec:Modelling2As}, $V$ plays a similar role to that of the Rabi frequency, however, we find that it instead parameterises the period of oscillation. As such, much like we see the average time between photons decrease in Fig.\ref{fig:g2s}, we expect the modification of $V$ to alter average photon spacings in a light period, and these results are shown in Fig.\ref{fig:photonstreams}. As seen in Fig.\ref{fig:photonstreams}{\color{blue}(d)}, by increasing $V$ we increase the length of dark periods, and this comes with a concurrent increase in the average time between photons. As predicted by the heatmap of $n_L$ in Fig.\ref{fig:heatmaps}, we observe no change in the average number of photons per light period, thus decreasing the visibility of each light period. The converse was true in {\color{blue}(e)}, where lowering $V$ lead to decreased dark and light period length, but increased the intensity of light periods. 
\begin{figure}[h!]
    \centering
    \includegraphics[width=\linewidth]{"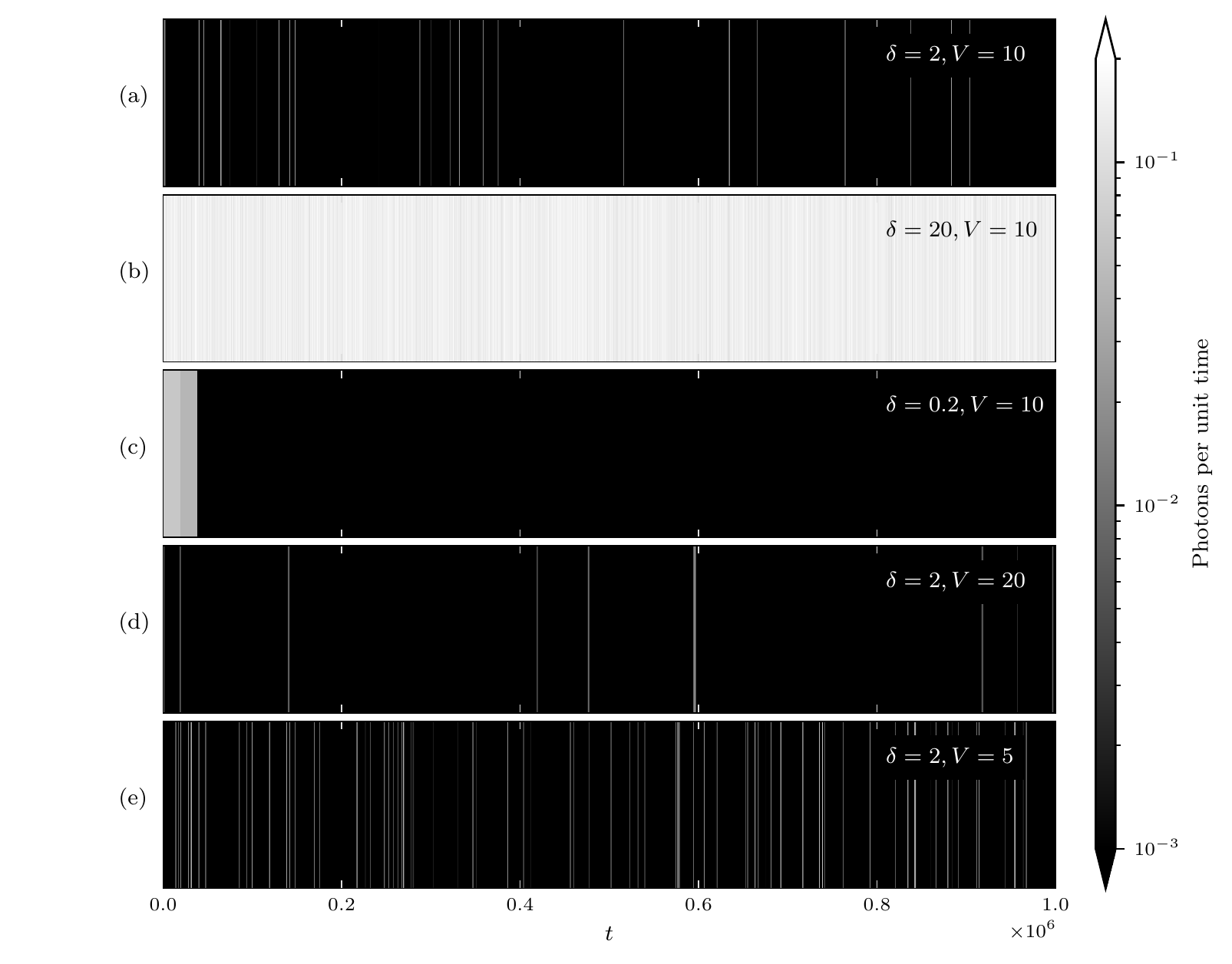"}
    \caption{Photon streams generated by binning jump times from simulations using configurations inset upper-right of each subplot into bin widths equal to each simulation's respective average light period length. Photons per unit time is then obtained by normalising the counts in each bin by the bin width. Simulated using system variables rescaled by factor $\Gamma$ giving: $\Gamma=1$, $\Gamma_{12}=1$ and $\Omega=5$, and simulation run for $t=1,000,000$.}
    \label{fig:photonstreams}
\end{figure}

Overall, we find good qualitative agreement with the heatmaps generated by Eq.(\ref{eq:TD}) for all period lengths and variable dependencies; for example being able to increase dark period length by either increasing $V$ or decreasing $\delta$, however, in order to demonstrate more readily the macroscopic jumps we generate photon stream {\color{blue}(f)} in Fig.\ref{fig:macrostream}. Here we see the ratio of light to dark period length better lends itself to easier observation of macroscopic jumps within the system, however, that this comes with a decrease in number of photons per unit time of nearly three orders of magnitude.

We finally note that Fig.\ref{fig:heatmaps} suggests it is possible to obtain dark states with arbitrarily high lifetimes in the limit of vanishing $\delta$, and this could make intuitive sense as we approach the maximally entangled antisymmetric state. However, for regimes with small differences in atomic transition frequencies it becomes difficult to determine the validity of the significantly extended dark periods predicted, due to the computational load required to simulate for the corresponding lengths of time.%in the case of stream (c), Eq.(\ref{eq:TD}) predicts dark periods of order $1\times10^{10}$, which is clearly not the case. 
%However, this fails in the configuration of stream (c), as we find that there is a significant amount of scope of nearly $\times 10^{7}$ for $T_{0}$, but that the integral in Eq.(\ref{eq:TD}) adds a vanishing contribution past $T_{0}$ compared to the possible values $T_{0}$ can take, greatly limiting the predictive ability of the approach.

\begin{figure}[h!]
    \centering
    \includegraphics[width=\linewidth]{"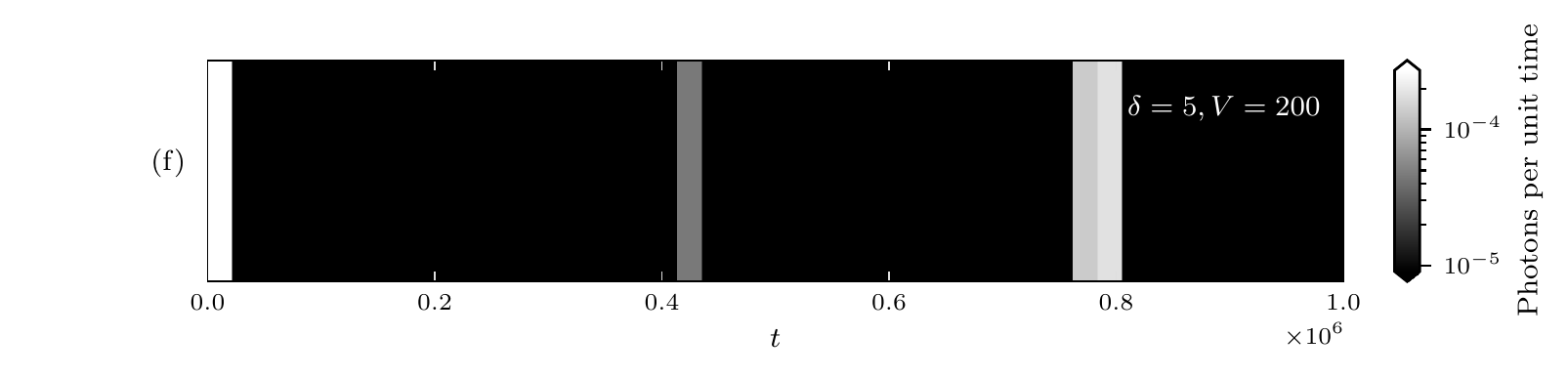"}
    \caption{Photon streams generated using the same methodology of Fig.\ref{fig:photonstreams}. Simulated using system variables rescaled by factor $\Gamma$ giving: $\Gamma=1$, $\Gamma_{12}=1$ and $\Omega=5$, and simulation run for $t=1,000,000$.}
    \label{fig:macrostream}
\end{figure}

\newpage
\section{Further Work \& Closing Remarks}
\label{sec:FurtherWork}
\vspace{-0.2cm}

%\subsection{Objective pure-state dynamical models}

Going forwards, whilst we have been able to find configurations of the arguments in QuTip's ODE solver that accurately simulate the longest timescales, further work is needed to fix the discrepancies in the shortest waiting times.
Regarding the macroscopic quantities in the final section of the report such as average dark period length, as shown by Hegerfeldt in Ref.\cite{DarkPeriods}, one can obtain analytical formulae for these quantities in terms of the system's parameters such as $\delta$ by obtaining the formulae of just the eigenvalues of $H_{\text{eff}}$, and hence be able to predict macroscopic characteristics directly from the system's configuration.
Unfortunately, analytical descriptions for systems such as the configuration of Fig.\ref{fig:UzmaEvo} are difficult to obtain as they inhabit an intermediary regime where nearly all parameters are of similar orders of magnitude, limiting the ability to approximate. Furthermore, the effective Hamiltonian of the two-atom dipole-dipole system is non-reducible, despite the excited state's minute population, due the the symmetric state's non-zero occupation post-jump, thus leading to a quartic characteristic polynomial of $H_{\text{eff}}$.
Despite this, further work in specific regimes, including investigations into the effect of driving strength on the phenomena displayed, may be useful in deepening our understanding of the nature of the dipole-dipole interaction in these two-atom systems.
In the context of the dipole-dipole induced collective effects in the two-atom system there is significant scope for further investigation, for example into the formation and enhancement of Mollow peaks, and researching the possibility of extending of the two-atom model to account for small fluctuations in position and orientation would allow for interesting work determining the stability of the macroscopic dark states formed, potentially providing results more easily examinable in a laboratory setting.

Due to the richness revealed when investigating the formation of macroscopic dark periods, we believe that the dipole-dipole interacting two-atom system alone is worthy of an entire research project of its own right. However, the demonstration of macroscopic jumps and the photon burst emission characteristics in the configuration of Section \ref{sec:DarkStates} show remarkable and intriguing parallels to Levy Flights and the evolutions of microbial populations. \cite{LevyFlights, Microbes, Dormancy} Furthermore, the quantum jump method may possess potential for modelling in econophysics, with the stochastically induced switching between different macroscopic phases of evolution that we have demonstrated holding promise in modelling transitions between periods of financial market behaviours.
\newline

\vspace{-0.1cm}
\label{sec:ClosingRemarks}

With regards to the deeper implications of the method, the brief discussions on measurement regimes were somewhat of a Pandora's box. Arguments as to whether Objective Pure State Models exist are still unresolved, with perhaps instead the detection scheme itself defining the behaviour of reality.\cite{OPDMs} This discussion is ongoing, with one side going as far as to assert that ``Quantum Jumps are more quantum than quantum diffusion'', and likely will continue for some time.\cite{VivaLaJump}
What is clear is that upon making the choice to employ standard photodetection methods, the quantum jump method is an eminently useful and flexible tool. We have shown that it allows for the concurrent demonstration and replication of known phenomena, elicits difficult discussions of the core implications of modern quantum mechanics, and by allowing for the simulation of individual trajectories and photo-emission events it has elucidated rich system dynamics with both interesting and potentially real-world applications, most definitely worthy of further investigation. 

%Despite the philosophical conundrums and the difficulties faced in tackling highly obtuse algebra, i

%\the\columnwidth is 15.75cm

\newpage
\vspace{0.3cm}
\begin{center}
    {
    \textbf{Acknowledgments}}
\end{center}
\label{subsubsec:Derivingg2}
\vspace{0.2cm}

%\begin{acknowledgments}
%\vspace{-0.3cm}
%\small
I would like to express my sincere gratitude to Professor Charles Adams for his continued, upbeat and ever insightful support throughout the duration of this research, and a special thanks goes to Dr Almut Beige, and to Nick Spong for their crucial insights and clarifications. \newline

%\end{acknowledgments}

}
%\newpage

\twocolumngrid{}

\onecolumngrid{}
\newpage
\appendix
\addtocontents{toc}{\protect\setcounter{tocdepth}{0}}
\section{}

\subsection{Electric Field vs Number Operator}
\label{app:Electricfieldvs}

As enabled by the quantum jump method, we investigate photon statistics in terms of a stream of discrete photon counts.  Thus, we must note that there is a minor difference in the formulation of correlation functions based on the electric field operator vs the photon number operator. By calculating $g^{(2)}(\tau)$ in terms of photon counts we are associating measurement events of the electric field with the detection of a photon, and experimentally this corresponds to interpreting the electronic signal registered by a photodetector as due to a photon that is localised. However, the number operator refers to the total number of photons in all space and hence is not strictly measureable.  With certain dimensions of real detectors one can justify this interpretation, but for simplicity this localisation of the number operator was reconciled by considering our detectors involved in photon statistics to be broadband and over all space. It is key to note that these broadband detectors involved in our calculation of photon statistics are different to those discussed in Section \ref{subsec:MeasurementTheory} on measurement theory. That is to say we simulate the system and the produced stream of photons is then ``sent" to our new idealised set of detectors to calculate the photon statistics. - 
%\newpage
L. Mandel and E. Wolf (1995) \textit{Optical Coherence and Quantum Op-
tics} Cambridge University Press

\newpage
\subsection{Matrix Representations}
\subsubsection{Single atom relaxing}

\noindent
\underline{System hamiltonian}:
\begin{equation}
    H_{S}=\frac{\hbar\omega}{2}\begin{pmatrix}
    1 & 0\\
    0 & -1
    \end{pmatrix},
\end{equation}
\underline{Dipole raising and lowering operators}:
\begin{equation}
    S^+=\begin{pmatrix}
    0 & 1\\
    0 & 0
    \end{pmatrix}, 
    \qquad     
    S^-=\begin{pmatrix}
    0 & 0\\
    1 & 0
    \end{pmatrix},
    \qquad
    S^+S^-=\begin{pmatrix}
    1 & 0\\
    0 & 0
    \end{pmatrix},
\end{equation}
\underline{Effective Hamiltonian}:
\begin{equation}
    H_{\text{eff}}=\frac{\hbar}{2}\begin{pmatrix}
    \omega-i\Gamma & 0\\
    0 & -\omega
    \end{pmatrix},
\end{equation}

\subsubsection{Driven single atom}
\noindent
\underline{System Hamiltonian}:
\begin{equation}
    H_{S} = H_{\text{opt}} =\frac{\hbar}{2}\begin{pmatrix}
        \Delta & \Omega\\[1mm]
        \Omega & -\Delta 
        \end{pmatrix}, \quad \Delta=\frac{1}{2}\left(\omega_{L} - \omega_{i} \right),
\end{equation}
\underline{Effective Hamiltonian}:
\begin{equation}
    H_{\text{eff}} =\frac{\hbar}{2}\begin{pmatrix}
        \Delta -i\Gamma & \Omega\\[1mm]
        \Omega & -\Delta 
        \end{pmatrix}, \quad \Delta=\frac{1}{2}\left(\omega_{L} - \omega_{i} \right),
\end{equation}

\subsubsection{Modelling two atoms in tensor product space}
\noindent
\underline{Hamiltonian 1}:
\begin{equation}
    H_{1\text{opt}} \otimes \mathds{1}_{2} =\frac{\hbar}{2}\begin{pmatrix}
        \Delta_1 & 0 & \Omega & 0\\[1mm]
        0 & \Delta_1 & 0 & \Omega\\[1mm]
        \Omega & 0 & -\Delta_1 & 0\\[1mm]
        0 & \Omega & 0 & -\Delta_1
        \end{pmatrix}, \quad \Delta_1=\frac{1}{2}\left(\omega_{L} - \omega_{1} \right),
\end{equation}
\underline{Hamiltonian 2}:
\begin{equation}
    \mathds{1}_{1} \otimes H_{2\text{opt}} =\frac{\hbar}{2}\begin{pmatrix}
        \Delta_2 & \Omega & 0 & 0\\[1mm]
        \Omega & -\Delta_2 & 0 & 0\\[1mm]
        0 & 0 & \Delta_2 & \Omega\\[1mm]
        0 & 0 & \Omega & -\Delta_2
        \end{pmatrix}, \quad \Delta_2=\frac{1}{2}\left(\omega_{L} - \omega_{2} \right),
\end{equation}
\underline{Dipole raising and lowering operators}:
\begin{equation}
    S_1^+  \equiv S_1^+ \otimes  \mathds{1}_{2}  =\begin{pmatrix}
        0 & 1 & 0 & 0\\
        0 & 0 & 0 & 0\\
        0 & 0 & 0 & 1\\
        0 & 0 & 0 & 0
        \end{pmatrix}, 
     \qquad
     S_2^+  \equiv\mathds{1}_{1} \otimes S_2^+   =\begin{pmatrix}
        0 & 0 & 1 & 0\\
        0 & 0 & 0 & 1\\
        0 & 0 & 0 & 0\\
        0 & 0 & 0 & 0
        \end{pmatrix}, 
\end{equation}
\underline{Dipole interaction Hamiltonian}:
\begin{equation}
        H_{dd} =\hbar \begin{pmatrix}
        0 & 0 & 0 & 0\\
        0 & 0 & V & 0\\
        0 & V & 0 & 0\\
        0 & 0 & 0 & 0
        \end{pmatrix}, 
\end{equation}
\underline{System Hamiltonian}:
\begin{equation}
    \begin{aligned}
        H_{S}=\hbar\begin{pmatrix}
        -\Delta & \Omega/2 & \Omega/2 & 0\\
        \Omega/2 & \delta/2 & V & \Omega/2 \\
        \Omega/2 & V & -\delta/2 & \Omega/2 \\
        0 & \Omega/2 & \Omega/2 & \Delta
        \end{pmatrix},
    \end{aligned}
    \quad\quad
    \begin{aligned}
        \Delta&=\omega_{L}-\frac{1}{2} \left( \omega_{1} + \omega_{2} \right) \\[1mm]
        \delta&=(\omega_{1}-\omega_{2}),
    \end{aligned}
\end{equation}

\subsubsection{Modelling two atoms in eigenbasis}

\noindent
\underline{Diagonalised Hamiltonian}:
\begin{equation}
    H_{\text{eig}}=\hbar\begin{pmatrix}
        -\Delta & 0 & 0 & 0\\
        0 & \lambda & 0 & 0 \\
        0 & 0 & -\lambda & 0 \\
        0 & 0 & 0 & \Delta
        \end{pmatrix}, \quad \lambda = \sqrt{\frac{\delta^{2}}{4} + V^{2}},
\end{equation}

\begin{equation}
    D=\begin{pmatrix}
        1 & 0 & 0 & 0\\
        0 & \alpha & \beta & 0 \\
        0 & \beta & -\alpha & 0 \\
        0 & 0 & 0 & 1
        \end{pmatrix}, \quad
    \alpha= \frac{1}{\left( 1 + \frac{\left(\frac{\delta}{2} - \lambda\right)^{2}}{V^{2}} \right)^{1/2}}
    ,\quad
    \beta= \frac{1}{\left( 1 + \frac{\left(\frac{\delta}{2} + \lambda\right)^{2}}{V^{2}} \right)^{1/2}},
\end{equation}
\underline{Dipole raising and lowering operators}:
\begin{equation}
        S_1^+ =\hbar \begin{pmatrix}
        0 & \beta & -\alpha & 0\\
        0 & 0 & 0 & \alpha\\
        0 & 0 & 0 & \beta\\
        0 & 0 & 0 & 0
        \end{pmatrix}, 
        \qquad\qquad
        S_2^+   = \begin{pmatrix}
        0 & \alpha & \beta & 0\\
        0 & 0 & 0 & \beta\\
        0 & 0 & 0 & -\alpha\\
        0 & 0 & 0 & 0
        \end{pmatrix}, 
\end{equation}

\underline{``Uzma'' operators}:
\begin{equation}
        U_s^+  =\frac{1}{\sqrt{2}} \begin{pmatrix}
        0 & \alpha + \beta & \beta - \alpha & 0\\
        0 & 0 & 0 & \alpha+ \beta\\
        0 & 0 & 0 & \beta-\alpha\\
        0 & 0 & 0 & 0
        \end{pmatrix}, 
\end{equation}
\begin{equation}
        U_a^+  =\frac{1}{\sqrt{2}} \begin{pmatrix}
        0 & \beta - \alpha  & -\beta - \alpha & 0\\
        0 & 0 & 0 & \alpha - \beta\\
        0 & 0 & 0 & \alpha + \beta\\
        0 & 0 & 0 & 0
        \end{pmatrix}, 
\end{equation}
\underline{System Hamiltonian}:
\begin{equation}
    H_{S}=\hbar\begin{pmatrix}
        -\Delta & \frac{\Omega(\alpha+\beta)}{2} & \frac{\Omega(\beta-\alpha)}{2} & 0\\
        \frac{\Omega(\alpha+\beta)}{2} & \lambda & 0 & \frac{\Omega(\alpha+\beta)}{2} \\
        \frac{\Omega(\beta-\alpha)}{2} & 0 & -\lambda & \frac{\Omega(\beta-\alpha)}{2} \\
        0 & \frac{\Omega(\alpha+\beta)}{2} & \frac{\Omega(\beta-\alpha)}{2} & \Delta
        \end{pmatrix}, \quad \lambda = \sqrt{\frac{\delta^{2}}{4} + V^{2}},
\end{equation}
\underline{Effective Hamiltonian}:
\begin{equation}
    H_{\text{eff}}=\hbar\begin{pmatrix}
        -\Delta -\frac{i\Gamma_s}{2}(\alpha^2 +\beta^2) & \frac{\Omega(\alpha+\beta)}{2} & \frac{\Omega(\beta-\alpha)}{2} & 0\\[1mm]
        \frac{\Omega(\alpha+\beta)}{2} & \lambda - \frac{i\Gamma_s}{4}(\alpha +\beta)^2 & \frac{i\Gamma_s}{4}(\alpha^2- \beta^2) & \frac{\Omega(\alpha+\beta)}{2} \\[1mm]
        \frac{\Omega(\beta-\alpha)}{2} & \frac{i\Gamma_s}{4}(\alpha^2- \beta^2) & -\lambda - \frac{i\Gamma_s}{4}(\alpha - \beta)^2 & \frac{\Omega(\beta-\alpha)}{2} \\[1mm]
        0 & \frac{\Omega(\alpha+\beta)}{2} & \frac{\Omega(\beta-\alpha)}{2} & \Delta
        \end{pmatrix}, \quad \lambda = \sqrt{\frac{\delta^{2}}{4} + V^{2}},
\end{equation}

\newpage
\subsection{Pairwise orthogonality}
\label{app:ortho}
To see this we construct the matrix $\Lambda$ of inner products of the right eigenstates, with $\Lambda_{ij}=\braket{\lambda_i|\lambda_j}$, and calculate the Frobenius norm of this lambda matrix minus the identity. As the diagonal entries of $\Lambda$ are equal to 1 by construction, this gives us a measure of orthogonality of the eigenstates. As seen in figure, we see that for all regions our parameter space this measure yeilds at most a 10e-4 difference between $\Lambda$ and the identity matrix.

\begin{figure}[h!]
    \centering
    \includegraphics[width=\linewidth]{"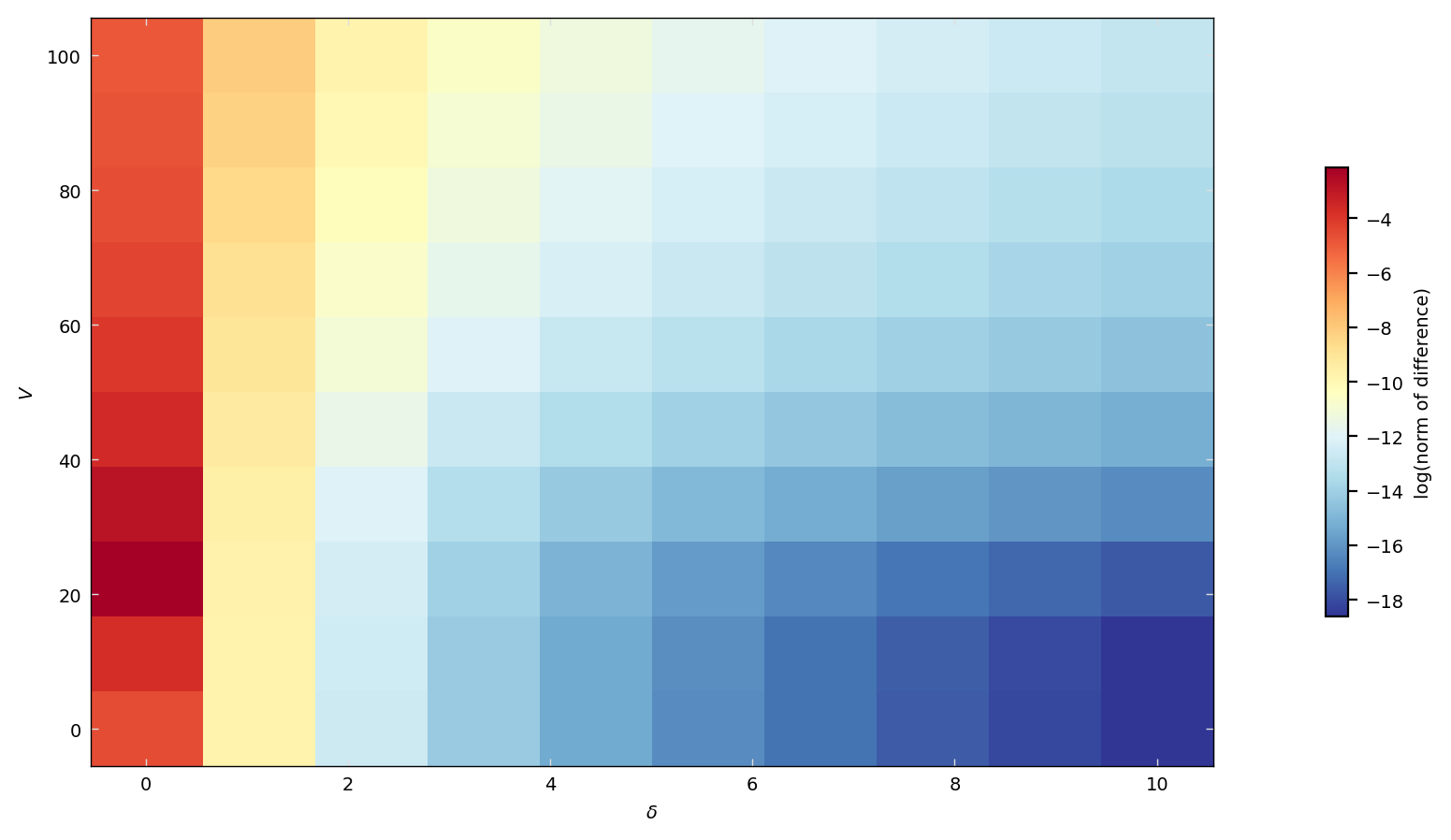"}
    %\caption{Caption}
    %\label{fig:my_label}
\end{figure}

\newpage
\subsection{Choosing $T_{apex}$}
\label{app:Tapex}

\begin{figure}[h!]
    \centering
    \includegraphics[width=\linewidth]{"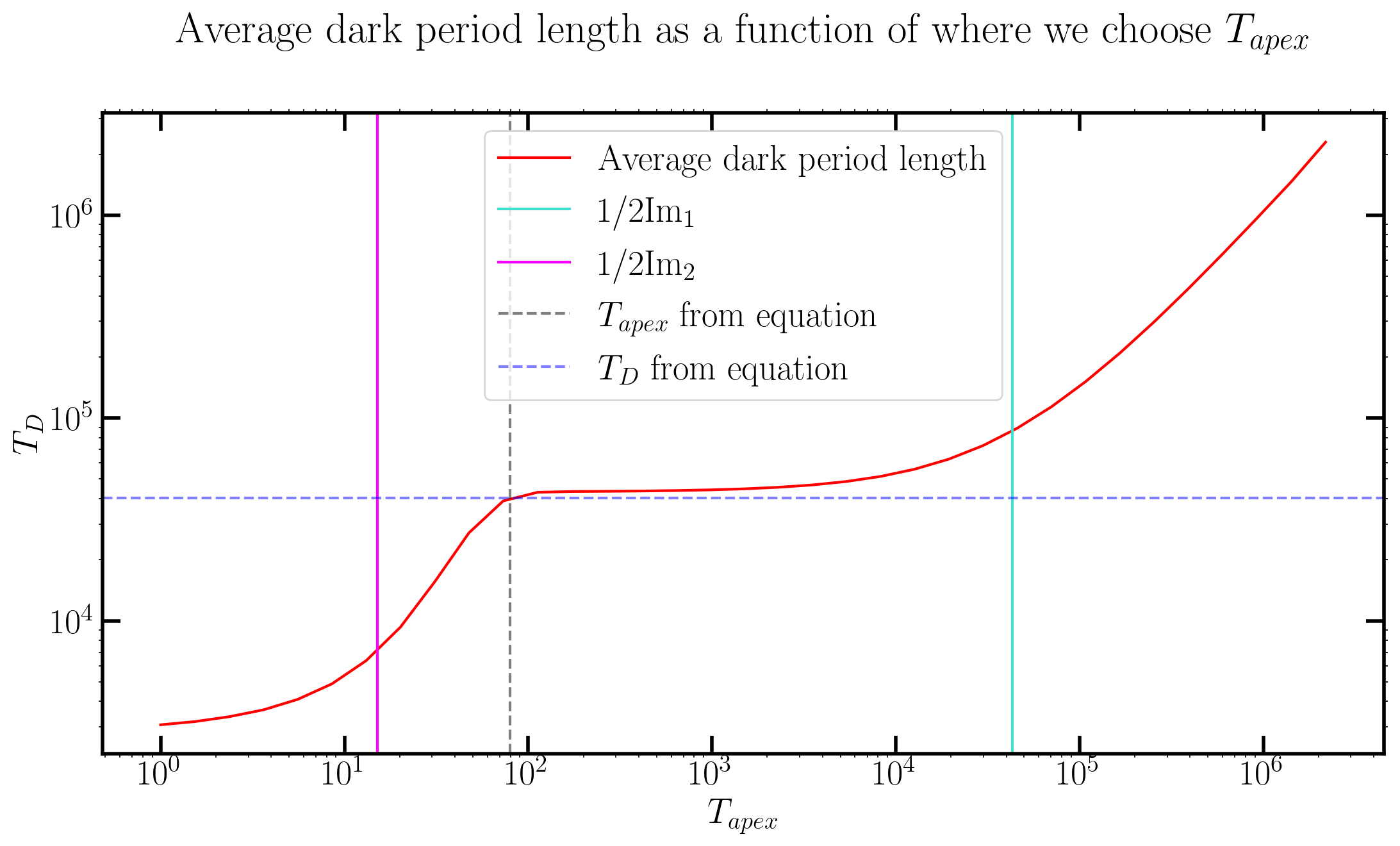"}
    %\caption{Caption}
    %\label{fig:my_label}
\end{figure}

\begin{figure}[h!]
    \centering
    \includegraphics[width=\linewidth]{"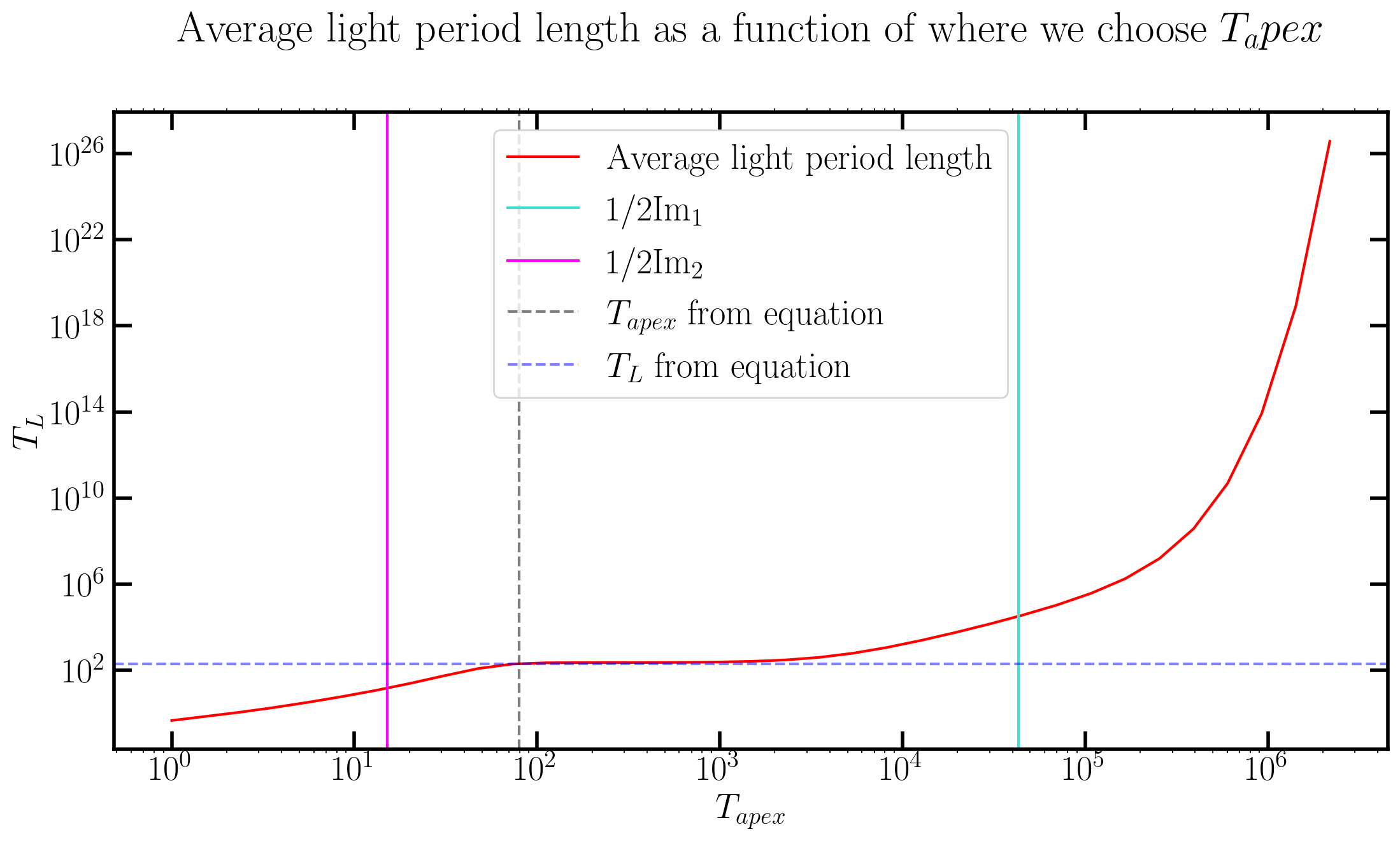"}
    %\caption{Caption}
    %\label{fig:my_label}
\end{figure}

\end{document}